\documentclass[pra,floatfix,twocolumn,superscriptaddress,showpacs,preprintnumbers,longbibliography,nofootinbib]{revtex4-1}
\usepackage{dcolumn}    
\usepackage{bm} 
\usepackage{graphicx}
\usepackage{amsmath}    
\usepackage{enumitem}
\usepackage{latexsym}
\usepackage{amsfonts}   
\usepackage{amssymb}
\usepackage{array}      
\usepackage{epsfig}
\usepackage{braket}
\usepackage{color}
\usepackage[colorlinks=true,linkcolor=blue,urlcolor=blue,citecolor=blue,pdfusetitle]{hyperref}
\usepackage{hyperref}
\usepackage[T1]{fontenc}
\usepackage[sc]{mathpazo}
\newcommand{\eq}{Eq.~}
\newcommand{\eqs}{Eqs.~}
\newcommand{\fig}{Fig.~}

\newcommand{\cf} {cf.~}

\newcommand{\rref} {Ref.~}
\newcommand{\rrefs} {Refs.~}

\begin{document}

\title{Mechanism of decoherence-free coupling between giant atoms}

\author{Angelo~Carollo}
\affiliation{Universit\`a  degli Studi di Palermo, Dipartimento di Fisica e Chimica -- Emilio Segr\`e, via Archirafi 36, I-90123 Palermo, Italy}
\affiliation{Radiophysics Department, National Research Lobachevsky State University of Nizhni Novgorod, 23 Gagarin Avenue, Nizhni Novgorod 603950, Russia}
\author{Dario~Cilluffo}
\affiliation{Universit\`a  degli Studi di Palermo, Dipartimento di Fisica e Chimica -- Emilio Segr\`e, via Archirafi 36, I-90123 Palermo, Italy}
\affiliation{NEST, Istituto Nanoscienze-CNR, Piazza S. Silvestro 12, 56127 Pisa, Italy}
\author{Francesco~Ciccarello}
\affiliation{Universit\`a  degli Studi di Palermo, Dipartimento di Fisica e Chimica -- Emilio Segr\`e, via Archirafi 36, I-90123 Palermo, Italy}
\affiliation{NEST, Istituto Nanoscienze-CNR, Piazza S. Silvestro 12, 56127 Pisa, Italy}

\begin{abstract}
Giant atoms are a new paradigm of quantum optics going beyond the usual local coupling. Building on this, a new type of decoherence-free (DF) many-body Hamiltonians was shown in a broadband waveguide. 
Here, these are incorporated in a general framework (not relying on master equations) and contrasted to dispersive DF Hamiltonians with normal atoms: the two schemes are shown to correspond to qualitatively different ways to match the same general condition for suppressing decoherence.
 Next, we map the giant atoms dynamics into a cascaded collision model (CM), providing an intuitive interpretation of the connection between non-trivial DF Hamiltonians and coupling points topology.
The braided configuration is shown to implement a scheme where a shuttling system subject to periodic phase kicks mediates a DF coupling between the atoms. From the viewpoint of CMs theory, this shows a collision model where ancillas effectively implement a dissipationless, maximally-entangling two-qubit gate on the system.
\end{abstract}

\maketitle

\section{Introduction}

Engineering decoherence-free (DF) Hamiltonians is a major task in the field of quantum technologies and many-body physics, with special regard to quantum optics implementations \cite{MolmerPRL99,BeigePRL00,KempePRA01,FacchiJPA2008,Shahmoon2013,Douglas2015b,PaulischNJP2016}. In particular, DF {\it mediated} Hamiltonians describe coherent interactions, typically between (pseudo) atoms or qubits which crosstalk via a quantum bus (usually some photonic environment) that yet does not introduce decoherence \cite{CohenAP}. In terms of the Lindblad master equation \cite{breuerFoundations2012}, 
this implies realizing a net second-order Hamiltonian that couple the atoms to one another, getting rid at once of the (usually present) dissipator term. Thereby one is left with an effective unitary dynamics of the atoms, where the environmental degrees of freedom are eliminated.

One of the typical strategies to achieve DF Hamiltonians is adiabatic elimination in the dispersive regime. In cavity QED, it is typically obtained by coupling a set of atoms far {off-resonantly} to cavity modes \cite{ZhengPRL2000,MajerNat07,IrishPRA08}. This gives rise to a separation of time scales such that incoherent second-order interactions average to zero, while coherent ones result in an effective Hamiltonian. An analogous working principle underpins DF Hamiltonians in structured photonic lattices \cite{Shahmoon2013,Douglas2015b,TudelaRMP18}, which are seeded by tuning the atomic frequency within a photonic band gap entailing an off-resonant interaction with all the lattice modes (this results in short-range, potentially tunable, inter-atomic couplings).

\begin{figure}
	\includegraphics[width=0.42\textwidth]{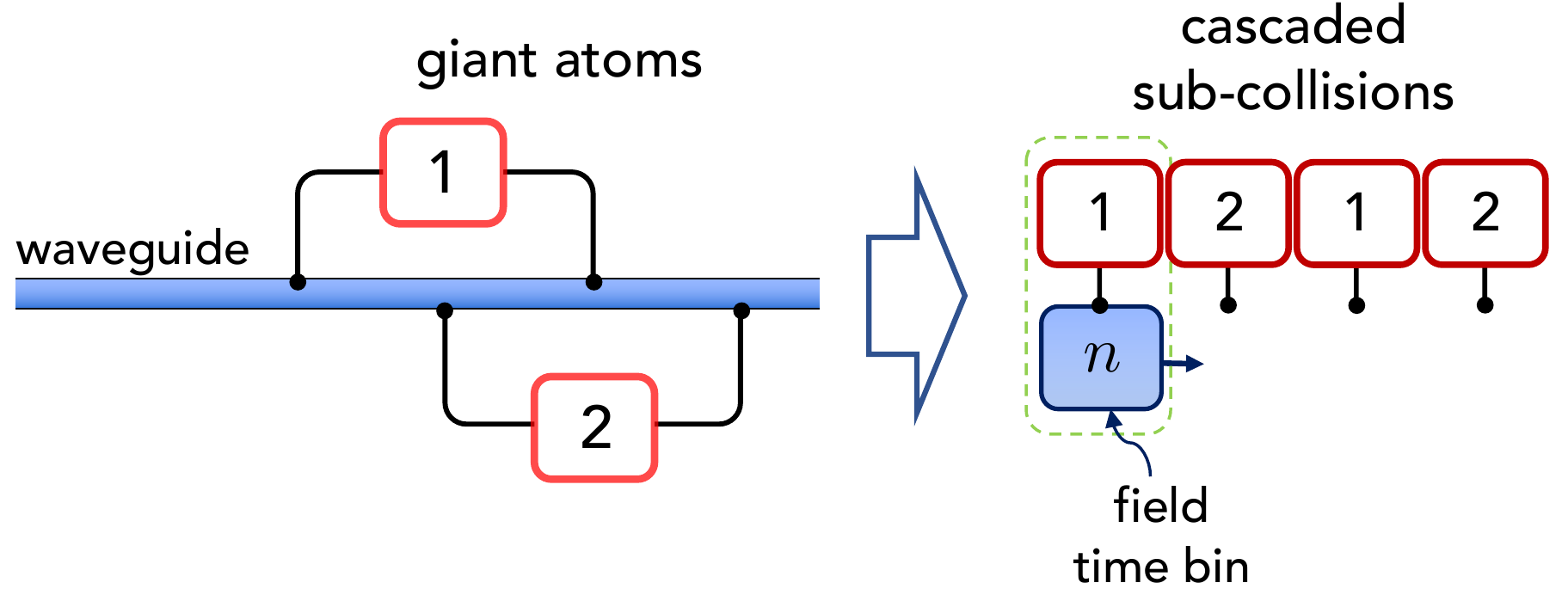}
	\caption{Two giant atoms, each coupled to a waveguide at two coupling points. The dynamics is mapped into cascaded sub-collisions, each involving one field time bin and one coupling point. }\label{sketch}
\end{figure}

Recently, a new class of DF Hamiltonians was predicted \cite{KockumPRL2018} and experimentally observed \cite{OliverGiant2019}, which employs giant atoms \cite{Kockum5years} in broadband waveguides. Giant atoms are a new playground of quantum optics \cite{FriskKockumPRA14,KockumPRA2017,Hammerer2019,AnderssonarXiv18,guimond2020unidirectional,guo2019oscillating,Wilson2020}, where the usual pointlike model of an emitter (normal atom) breaks down. In contrast, as sketched in \fig\ref{sketch}, a giant atom (typically an artificial two-level system) couples to the field at a discrete set of distinct coupling points (an alternative implementation is an atom in front of a mirror \cite{WitthautNJP10,FangNJP18,HoiNatPhy15}). By appropriate engineering, the distance between coupling points can be made several wavelengths long. This introduces tunable phase shifts, yielding interference effects unattainable with normal atoms. Notably, these can be harnessed in particular to suppress dissipative interactions, giving rise to DF Hamiltonians \cite{KockumPRL2018,Hammerer2019}. As a distinctive feature of giant atoms, such Hamiltonians can happen to be trivial (i.e., identically zero or with null coupling terms) depending on the coupling points topology. For two giant atoms, for instance, only one out of the three possible topologies leads to non-trival DF Hamiltonians \cite{KockumPRL2018,OliverGiant2019}.

So far occurrence of giant atoms DF Hamiltonians was mostly investigated through the explicit master equation of the atoms \cite{KockumPRL2018} derived via the SLH formalism \cite{CombesAdvPhyX17}, in turn related to the input-output theory for waveguide-QED setups \cite{LalumierePRA13}. This master equation is a rather involved object featuring both local and non-local dissipative terms, based on which the essential mechanism behind emergence of DF Hamiltonians is not straightforwardly interpreted. In the case of the aforementioned DF dispersive Hamiltonians, however, a derivation solely based on average Hamiltonian theory \cite{ernst1987principles,ChuangRMP05,JamesJCan07,JamesPRA10,FalciPRA16} (thus not relying on master equations) is possible \cite{IrishPRA08}. One thus naturally asks whether a similar picture can be defined for giant atoms and used to study conditions for occurrence of DF Hamiltonians. 

With the above motivations, in this work we consider a general framework for deriving DF Hamiltonians in the spirit of average Hamiltonian theory. The main condition (``DF condition") is to arrange for an interaction Hamiltonian averaging to zero over a coarse-grained time scale (in the interaction picture). Dispersive schemes and giant atoms in a broadband waveguide are compared and shown to be different ways to match the DF condition. This occurs through destructive interference of a continuum of phase factors defining the interaction Hamiltonian in the former case and only a discrete, possibly small, number in the latter. We next focus on giant atoms schemes, reviewing their general description through a collision-model picture and connecting it to the DF-Hamiltonians framework. It is then shown that each collective collision with all the atoms can be decomposed as a cascade of ordered subcollisions (see \fig\ref{sketch}), each involving a single coupling point. Topologies yielding a zero effective Hamiltonian (such as serial and nested configurations) correspond to subcollisions combined with their time-reversed analogues so as to produce a zero net evolution. Instead, non-zero DF Hamiltonians are seeded for topologies (such as the braided) that feature combinations of subcollisions and their time-reversed versions not leading to an overall identity evolution. 

Moreover, we show that the cascaded-collision picture allows us to map giant atoms dynamics into a mediator (embodied by a field time bin) shuttling between the atoms and subject to periodic phase kicks. The equivalent quantum circuit, a sequence of parametric iSWAP and local phase gates, is presented.

This paper starts in Sec.~\ref{general} by developing a general framework for occurrence of DF Hamiltonians based on the Magnus expansion of the joint propagator in each time interval. We next review in Sec.~\ref{sec-ae} how the necessary requirements for having a DF Hamiltonian are fulfilled in the case of standard dispersive Hamiltonians: this provides an illustration of the theory of Sec.~\ref{general} in a familiar setup with which DF Hamiltonians via giant atoms (our main focus) will be profitably compared. In Sec.~\ref{sec-giant}, giants atoms coupled to a broadband waveguide are introduced, showing how they provide a different way to match the DF condition compared to dispersive schemes. In Sec.~\ref{sec-CP}, we review how the joint dynamics of giant atoms and field can be described through an average-Hamiltonian approach, decomposing into a sequence of elementary unitaries in each of which the atoms jointly collide with a field time bin. Sec.~\ref{CM-def} shows how each collision in turn can be decomposed into cascaded subcollisions, one for each coupling point. This highlights the physical origin of the effective Hamiltonian so as to link it to the coupling points topology, a task carried out in Sec.~\ref{sec-mec}. In Sec.~\ref{sec-qubit}, we show that the giant-atoms setup can be seen as an implementation of a scheme where a shuttling qubit, subject to periodic phase gates, mediates an indirect DF coupling between the atoms, and the equivalent quantum circuit is given. The theory developed in Sections \ref{CM-def} and \ref{sec-mec} is extended to a bidirectional (generally chiral) waveguide in Sec.~\ref{sec-bid}, and to more than two coupling points in Sec.~\ref{sec-multi}. Finally, we draw our conclusions in Sec.~\ref{sec-conc}.

\section{General scheme for decoherence-free Hamiltonians}\label{general}

Consider an unspecified quantum system $S$ coupled to a quantum  environment $E$. The Hamiltonian reads
\begin{equation}
H= H_S+H_E+ V\,,\label{Htot}
\end{equation}
with $H_S$ ($H_E$) the free Hamiltonian of $S$ ($E$) and $V$ their interaction Hamiltonian. In the interaction picture with respect to $ H_0=H_S+H_E$, the joint state $\sigma_t$ evolves as 
\begin{equation}
\dot {\sigma}_t=-i\,  [V_t,\sigma_t]
\end{equation}
with $V_t=e^{i H_0 t}\,V\, e^{-i H_0 t}$.
Thus at time $t$, 
\begin{equation}
\sigma_t= {\cal U}_t\, \sigma_0 \,{\cal U}_t^\dag
\end{equation}
with the propagator $ {\cal U}_t$ given by $ {\cal U}_t= {\mathcal{T}} \,\exp[{-i\int_{t_{0}}^{t} ds\,  V(s)}]$, 
where ${\cal T}$ is the usual time ordering operator. 

Consider now a mesh of the time axis defined by $t_n=n\Delta t$ with $n=0,1,...$ and $\Delta t$ the time step, in terms of which the propagator can be decomposed as
\begin{equation}
{\cal U}_t=\prod_{n=1}^{[t/\Delta t]}  U_n\,\,\,\,\,{\rm with}\,\,\, U_n= {\mathcal{T}} \,e^{-i\int_{t_{n-1}}^{t_n} ds\,  V(s)}\,.\label{Ut}
\end{equation}
Now, if $\Delta t$ is short enough compared to the characteristic time of interaction, applying the Magnus expansion \cite{Magnus1954} each unitary $ U_n$ can be approximated to second order as
\begin{align}
 U_n\simeq \openone -i \,( \overline V_n  +  {\cal H}_{n} )\, \Delta t -\tfrac{1}{2} \overline V_n^2 \Delta t^2\label{Un-app}\,
\end{align}
with $\openone$ the identity operator and
\begin{eqnarray}
\overline{V}_n &=&\tfrac{1}{\Delta t}\!\int_{t_{n-1}}^{t_n} \!ds \, V_s\,,\label{Vn}\\
 {\cal H}_{n} &=&-\tfrac{i}{2\Delta t}\!\int_{t_{n-1}}^{t_n} \!ds \int_{t_{n-1}}^{s} \!ds' \,[ V_s, V_{s'}]\,.\label{Hn}
\end{eqnarray}
The averaged interaction $\overline{V}_n$ and Hamiltonian $ {\cal H}_{n}$, respectively of first and second order in the coupling strength, are the two central quantities to consider for implementing DF Hamiltonians. In sketchy terms, one seeks to fulfill $\overline{V}_n=0$ (henceforth referred to as the ``DF condition") in a way that ${\cal H}_{n}$ yields (upon partial trace) a dissipationless effective Hamiltonian of $S$, $H_{\rm eff}$. This is formalized in detail in the following. 

Let $\sigma_{n}$ be the joint $S$-$E$ state at time $t_n$ and $\rho_{n}={\rm Tr}_E\{\sigma_{n}\}$ the reduced state of the system at the same time. We will consider a coarse-grained time scale defined by $\Delta t$ short enough that \eqref{Un-app} holds. In the corresponding continuous-time limit, $t_{n}\rightarrow t$, $\sigma_{n}\rightarrow \sigma_t$, $\Delta\sigma_n/\Delta t\rightarrow \dot{\sigma}_t$ where we set $\Delta\sigma_n=\sigma_{n}-\sigma_{n-1}$ (analogously for $\rho_n$).

We also define 
\begin{eqnarray}
\langle  {\cal H}_{n} \rangle_{\rho_0}&=&{\rm Tr}_S\left\{{\cal H}_{n}\, \rho_0\otimes \openone_E\right\}\,,\label{Hn0}\\
H_{\rm eff}&=&{\rm Tr}_E\left\{{\cal H}_{n} \, \openone_S\otimes\rho_E\right\}\,\label{Heff}
\end{eqnarray}
with ${\rm Tr}_{S(E)}\{\}$ the partial trace over $S$ ($E$). These are effective Hamiltonians on $E$ and $S$, respectively. When $S$ is multipartite, in particular, $H_{\rm eff}$ will generally feature mutual couplings between subsystems of $S$.

The following property holds. 
\\
\\
\noindent {\bf Property.} {\it  Let the system and environment be initially in the uncorrelated state $\rho_0 \otimes \rho_E$ with $\rho_0$ ($\rho_E$) the initial state of the system (environment). If}
\begin{equation}
\overline V_n =0 \label{cond}
\end{equation}
{\it in each time interval $[t_{n-1},t_n]$, and
\begin{equation}
[ {\cal H}_{n},\openone_S\otimes\rho_E]=0 \label{cond2}
\end{equation}	
then in the continuous-time limit}
\begin{equation}
\dot \rho= -i\,[ H_{\rm eff}, \rho]\,.\label{evol}  %
\end{equation}

This embodies a rather general working principle for realizing DF effective Hamiltonians: conditions \eqref{cond} and \eqref{cond2} entail a {\it unitary} reduced dynamics of $S$ generated by the effective Hamiltonian $H_{\rm eff}$. Among \eqref{cond} and \eqref{cond2}, the former (DF condition) is the most relevant: it means that the interaction Hamiltonian $V_t$ {\it averages} to zero over the coarse-grained time scale $\Delta t$.

The above property is easily shown (see Appendix \ref{app-proof}), from which in particular it turns out that $\sigma_n=\rho_n \otimes \rho_E$ namely $E$ remains in its initial state, uncorrelated with $S$. 

A typical case where \eqref{cond2} occurs is when ${\cal H}_n$ acts trivially on $E$, then \eqref{cond2} is matched for any $\rho_E$ and ${\cal H}_n\equiv H_{\rm eff}$ (this happens with giant atoms as we will see). Another instance is when $S$ is a two-level system and $E$ a harmonic oscillator with ${\cal H}_n\sim\! \sigma_z\, b^\dag b$ (dispersive regime of the Jaynes-Cummings model \cite{harocheExploring2006}; see next section). Then \eqref{cond2} holds when $\rho_E$ is any mixture of Fock states.

Note that the above framework, alongside related approaches \cite{JamesJCan07,JamesPRA10,SorensenPRA11}, bypasses any direct use of master equations or the Born-Markov approximation, being instead mostly based on propagators and Hamiltonians. 

\section{Dispersive Hamiltonians}\label{sec-ae}

A longstanding way for matching condition \eqref{cond} in quantum optics is coupling atoms to a single- or multi-mode photonic environment dispersively, i.e., off-resonantly. A standard model to illustrate this is a set of identical two-level atoms of frequency $\omega_{0}$ and ground (excited) state $\ket{g}$ ($\ket{e}$) weakly coupled to a bosonic field. The atoms play the role of system $S$ and the field of environment $E$. Their free Hamiltonians read
\begin{equation}
H_S=\omega_{0}\sum_j  \sigma_{j}^\dag \sigma_{j}\,,\,\,H_E=\sum_k \omega_k\, b_k^\dag b_k \,,\label{free}
\end{equation}
while the interaction Hamiltonian in the rotating-wave approximation is given by
\begin{equation}
V= \sum_{j,k}   g_{jk}\, \sigma_{j}b_k^\dag +{\rm H.c.}\,\label{V}
\end{equation}
with coupling strength $g_{jk}$ generally complex, $\sigma_{j}=\ket{g}_j\!\bra{e}$ and $b_k$ bosonic ladder operators of the field (here $k$ labels the field normal modes and in general could comprise both discrete and continuous indexes).

In the interaction picture, $V$ turns into
\begin{equation}
V_t=\sum_j \sum_k  g_{jk} \,\sigma_{j}b_k^\dag \,e^{i  \Delta_{k} t}+{\rm H.c.}\,\label{Vt-ae}
\end{equation}
with $\Delta_{k}=\omega_k-\omega_0$ the detuning between mode $k$ and the atomic transition frequency $\omega_{0}$.

Consider now the off-resonance regime such that the detunings are all much larger than the typical order of magnitude of the interaction, which is generally expressed as ${\rm min}_k |\Delta_k|\gg {\rm max}_{jk} |g_{jk}|$. Then one can choose a coarse-grained time scale $\Delta t$ such that
\begin{equation}
\max_k |\Delta_k|^{-1}\ \ll \Delta t \ll \min_{jk} |g_{jk}|^{-1}\,.\label{sep}
\end{equation}
Accordingly, 
\begin{equation}
\int_{t_{n-1}}^{t_n}\! dt \,e^{\pm i \Delta_{k} t} \simeq 0\,\,\,{\rm for\,\,any}\,\,k\,,\label{ae-app}
\end{equation}
hence \eqref{Vt-ae} averages to zero in each time interval $[t_{n-1},t_n]$ of length $\Delta t$ so as to fulfill the DF condition \eqref{cond}. The Hamiltonian term \eqref{Hn} is given by
\begin{align}
{\cal H}_n=\!&-\sum_{j,j'} \sum_{k} \tfrac{g_{j,k} g_{j',k}^*}{2\Delta _k} \sigma_{j'}^\dag\sigma_{j}+{\rm H.c.}+\sum_{j} \sum_{k,k'}{}^{'} \,\tfrac{g_{j,k}\,g_{j,k'}^*}{\Delta _{k}} \sigma_{jz} b_{k}^\dag b_{k'}\,\nonumber
\end{align}
with the primed sum running  over all $k,k'$ such that $\omega_{k}=\omega_{k'}$. For a single atom and only one field mode, this reduces to the interaction Hamiltonian $\sim \!\sigma_z b^\dag b$ arising in the dispersive regime of the Jaynes-Cummings model \cite{HarocheRaimondBook}.
\begin{figure}
	\includegraphics[width=0.22\textwidth]{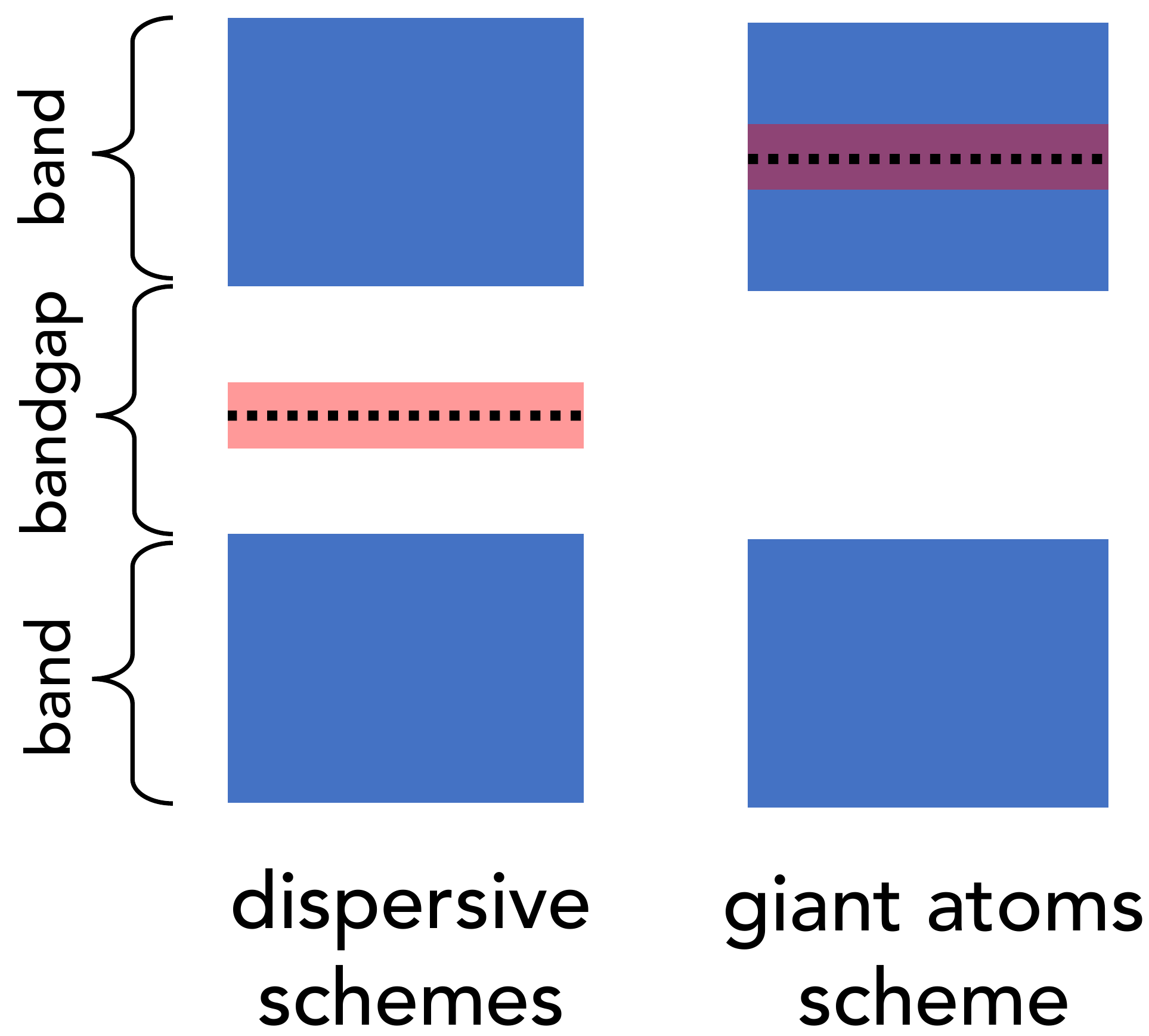}
	\caption{Frequency regimes for realizing DF Hamiltonians. The photonic environment typically features frequency bands separated by bandgaps. The dashed line marks the atomic frequency $\omega_{0}$, which couples resonantly with a bandwidth of modes (in red) of the order of the atom-field coupling rate.  Dispersive schemes (left panel) work off-resonance by tuning atoms far off-resonance from lattice bands. Instead, DF Hamiltonians with giant atoms (on the right) operate well within a photonic band, which can thus be approximated as infinite.}\label{band}
\end{figure}

Choosing $\rho_E=|0 \rangle \langle 0|$, with $\ket{0}$ the field vacuum state, condition \eqref{cond2} is fulfilled, hence we get the DF effective Hamiltonian [\cf\eq\eqref{Heff}]
\begin{equation}
H_{\rm eff}=-\sum_{j,j'} \sum_{k} \!\tfrac{g_{j,k} g_{j',k}^*}{2\Delta _k} \sigma_{j'}^\dag\sigma_{j}+{\rm H.c.}\,
\end{equation}
featuring atom-atom couplings.

\section{Giant atoms in a broadband waveguide}\label{sec-giant}

A standard way to realize the scheme in the previous section is to couple the atoms to a photonic lattice and tune $\omega_0$ far from any band (see \fig\ref{band}). The atoms then interact with the photonic environment far {off-resonantly}, which results in the separation of time scales \eqref{sep}. 

\begin{figure}
	\includegraphics[width=0.4\textwidth]{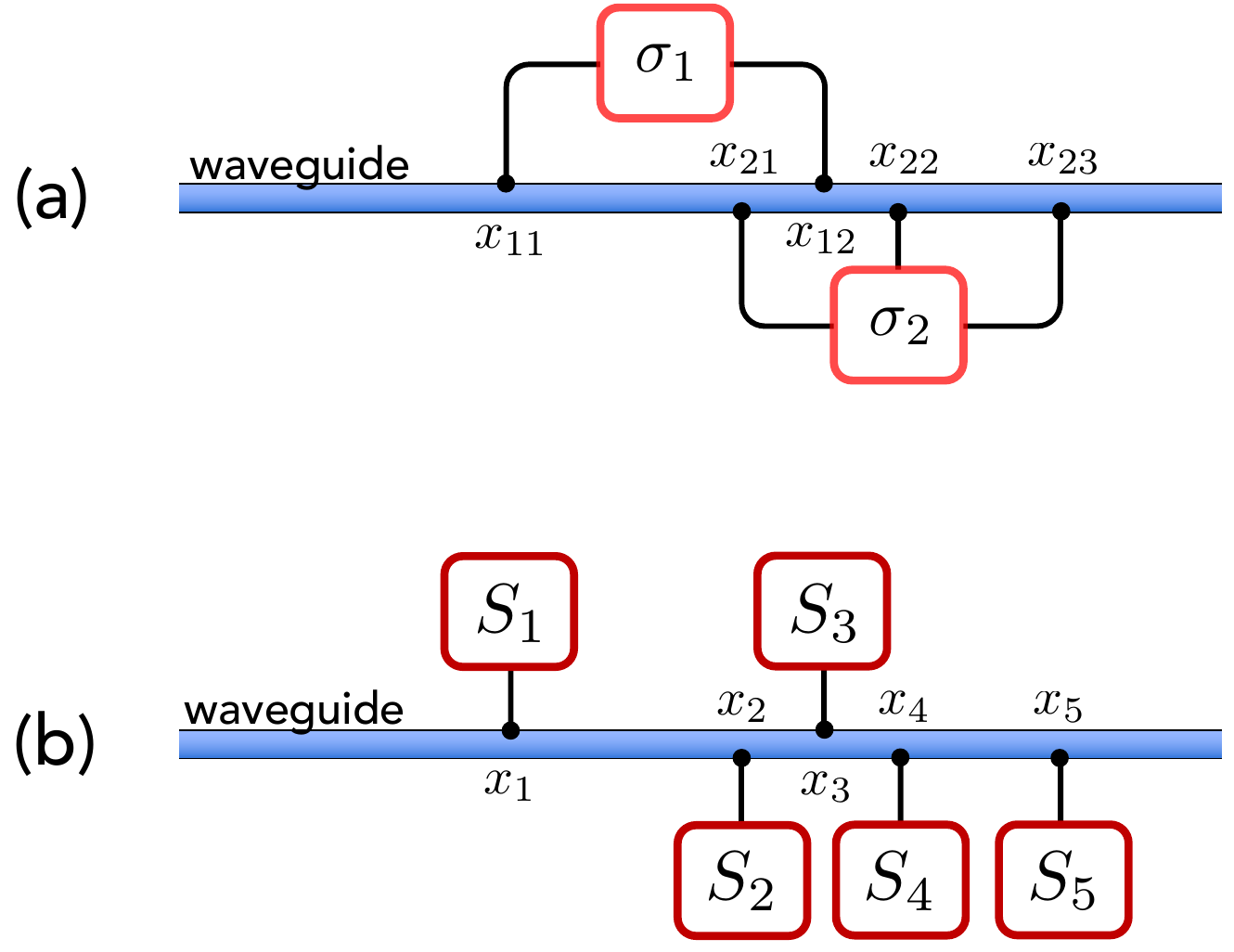}
	\caption{Giant atoms coupled to a waveguide. (a): Instance with two atoms (1 and 2) having ${\cal N}_1=2$ and ${\cal N}_2=3$ coupling points, each of coordinate $x_{j\ell}$ with $x_{j1}<x_{j2}<...\,$. (b): For each coupling point $\nu=1,...,{\cal N}$ ($\nu$ growing from left to right) we define an atomic operator $S_{\nu}$ [\cf \eq\eqref{Snu}] as the $\sigma_j$ of the corresponding atom times the coupling-point phase factor $e^{-i \varphi_{j\ell}}$. E.g., $S_1=\sigma_1 e^{-i \varphi_{11}}$, $S_4=\sigma_2 e^{-i \varphi_{22}}$ with $\varphi_{j\ell}=k_0 x_{j\ell}$. Note that $S_\nu$ generally does not commute with $S_{\nu'\neq \nu}^\dag$. Atomic operators $S'_{\nu}$ are defined analogously, except that the phase changes sign, $\varphi_{j\ell}\rightarrow -\varphi_{j\ell}$.}\label{labels}
\end{figure}

Instead, decoherence-free Hamiltonians via giant atoms work in the regime in which the atomic frequency $\omega_{0}$ is well within a photonic band which can thus be approximated as infinite (see \fig\ref{band}). This is possible due to {\it non-local} coupling (the hallmark of giant atoms) as will become clear later.
Thus consider a set of giant two-level atoms weakly coupled to a one-dimensional waveguide \cite{RoyRMP17,LiaoPhyScr16,GuarXiv17} with $\omega_0$ inside a band of the waveguide field. The free atomic Hamiltonians of $S$ and $E$ are still given by \eq\eqref{free}, where (compared to the general case in the previous section) $k$ is now intended as the wavevector. 
The $j$th atom is coupled to the waveguide at ${\cal N}_j$ distinct coupling points [see \fig\ref{labels}(a)], the coordinate of each being $x_{j\ell}$ with $\ell=1,..., {\cal N}_j$ (such that $x_{j1}<x_{j2}<...$). 
Accordingly, the interaction Hamiltonian in the interaction picture now reads 
\begin{equation}
V_t= \sum_{j,\ell}\sum_k  g_{j\ell k}\, \sigma_{j}b_k^\dag e^{i \Delta_k t}+{\rm H.c.}
\end{equation}
with $g_{j\ell k}$ the coupling strength to mode $k$ of the $\ell$th coupling point of atom $j$. 
Unlike the previous section, resonant modes $k\simeq \pm k_0$ (with $\omega_{k_0}=\omega_0$) will now dominate, thus \eqref{ae-app} does not hold. 

The coupling strengths more explicitly read $g_{j\ell k}= g_k \,e^{-i k x_{j\ell}}$, where taking advantage of weak coupling we can approximate $g_k\simeq g_{k_0}$
(which thus become $k$-independent). Accordingly, we can write
\begin{equation}
g_{j\ell k}= g_{k_0} \,e^{-i k_0 x_{j\ell}}e^{-i (k-k_0) x_{j\ell}}\,,
\end{equation}
where (for the sake of argument) we are assuming for now a unidirectional field.
Plugging into $V_t$, we get
\begin{align}
V_t=\,& g_{k_0}\sum_{j,\ell}e^{-i \varphi_{j\ell}}\sigma_{j}\sum_k \, e^{-i (\Delta_k t-k x_{j\ell})}b_k^\dag+\,{\rm H.c.}\,,\label{Vt-g}
\end{align}
where we defined the coupling point phases
\begin{equation}
\varphi_{j\ell}=k_0\, x_{j  \ell}\label{phase}
\end{equation}
and performed the variable change $k\rightarrow k- k_0$ (wave vector measured from $k_0$). Consistently with the weak-coupling regime, we can linearize the photonic dispersion law around the atomic frequency as $\omega_k\simeq \omega_{0}+ v k$ with $v$ the photon velocity. Using $v$, the coupling points coordinates can be expressed in the time domain as $\tau_{j\ell}=x_{j\ell}/v$.
Thereby, \eqref{Vt-g} becomes
\begin{equation}
V_t= g_{k_0}\sum_{j,\ell}e^{-i \varphi_{j\ell}}\sigma_{j}\sum_k \, e^{-i \omega_k (t-\tau_{j\ell})}\,b_k^\dag+\,{\rm H.c.}\label{Vt-g-2}\,.
\end{equation}

Averaging \eqref{Vt-g-2} over a time interval [$t_{n-1},t_n]$ yields
\begin{equation}
\overline V_n=g_{k_0}\sum_{j}\left(\sum_{\ell}e^{-i \varphi_{j\ell}}\right)\sigma_{j}\int_{t_{n-1}}^{t_n}\!ds\sum_k \, e^{-i \omega_k (s-\tau_{j\ell})}\,b_k^\dag+\,{\rm H.c.}\label{overV}
\end{equation}
(we have also split the sum over $j$ and $\ell$).
Note that $\tau_{j'\ell'}-\tau_{j\ell}$ is the time delay taken by light to travel from the $\ell$th coupling point of atom $j$ to the $\ell'$th coupling point of $j'$. If all these time delays are negligible compared to $\Delta t$, then \eqref{overV} can be approximated as
\begin{equation}
\overline V_n\simeq  g_{k_0}\sum_{j}\left(\sum_{\ell}e^{-i \varphi_{j\ell}}\right)\sigma_{j}\int_{t_{n-1}}^{t_n}\!ds\sum_k  e^{-i \omega_k s}\,b_k^\dag+\,{\rm H.c.}
\end{equation}
Now, the key point is that each atomic operator $\sigma_{j}$ comes with a pre-factor $\sum_{\ell}e^{i \varphi_{j\ell}}$, which -- due to non-local coupling -- can vanish for all atoms at the same time. This occurs when the coupling point phases are adjusted so as to match the condition
\begin{equation}
\sum_{\ell=1}^{{\cal N}_j}e^{-i \varphi_{j\ell}}=0\,\,\,\,{\rm for}\,\,{\rm any}\,\,j\,,\label{cond-g2}
\end{equation}
which is the DF condition \eqref{cond} for giant atoms.
Note that this cannot be satisfied by normal atoms: {\it each} atom must have at least {\it two} coupling points (${\cal N}_j\ge 2$). 

It is interesting to compare \eqref{cond-g2} with \eqref{ae-app}. Each can be seen as a destructive interference condition, involving a continuum of phase factors in the dispersive scheme but only a discrete, possibly small, number in the scheme with giant atoms.

\section{Giant atoms dynamics: average-Hamiltonian description}\label{sec-CP}

The previous section showed how the DF condition $\overline V_n=0$ is realized with giant atoms in comparison with off-resonance schemes. When it comes to giant atoms, depending on the topology of coupling points, the DF condition can result in a vanishing (thus trivial) $H_{\rm eff}$ \cite{nota-trivial}. Clarifying the requisites for obtaining a non-trivial effective Hamiltonian, and especially the related physical interpretation, is a major goal of this work. Prior to this, however, we reformulate the microscopic model with giant atoms in terms of time mode operators of the field (generalizing at once to a bidirectional, generally chiral, waveguide), this being the theoretical basis for the mapping of the dynamics into a cascaded collision model that will be discussed in the next sections.
\subsection{DF condition}
To begin with, as shown in \fig\ref{labels}(b), it is convenient to introduce a single index $\nu=1,...,{\cal N}$ labeling all the coupling points from left to right, where ${\cal N}=\prod_j {\cal N}_j$ is the total number of coupling points. For each coupling point so indexed, we define atomic operators depending on the corresponding pair $(j,\ell)$ as 
\begin{eqnarray}
S_{\nu}&=&\sigma_{j} \,e^{-i \varphi_{j\ell}}\,,\,\,S'_{\nu}=\sigma_{j} \,e^{i \varphi_{j\ell}}\,, \label{Snu}
\end{eqnarray}
For instance, in the case of \fig\ref{labels}, $S_3= \sigma_1 e^{-i \varphi_{12}}$ and $S'_3= \sigma_1 e^{i \varphi_{12}}$ with $\varphi_{12}=k_0 x_{12}$ (see Eq.~\eqref{phase}). Note that $S_\nu$ generally does not commute with $S_{\nu'\neq \nu}^\dag$ (e.g., in \fig\ref{labels}, $[S_2,S_5^\dag]\neq 0$).

Next, we come back to \eqref{Vt-g-2} and extend it to a bidirectional waveguide using the newly introduced operators as (see \rref\cite{CP} for more details)
\begin{align}
V_t=\,& g_{k_0}\sum_\nu S_\nu \sum_k \, e^{-i \omega_k (t-\tau_{\nu})}\,b_k^\dag\nonumber\\
&+g_{-k_0}\sum_\nu S'_\nu\sum_k   \,e^{i \omega_k (t+\tau_{\nu})}\,{b'}_k^\dag +\,{\rm H.c.}\,\label{Vt-g-3}
\end{align}
with ladder operators $b_k$ ($b'_k$) now corresponding to right-going (left-going) modes [in the first (second) sum $k$ is measured from $k_0$ ($-k_0$)].

In the limit in which the field becomes a continuum of modes, $V_t$ can be expressed in the form \cite{CP}
\begin{equation}
V_t=\sqrt{\gamma}\sum_{\nu}\,S_{\nu} \,{b}^\dag_{t-\tau_\nu}+\sqrt{\gamma'}\,\sum_{\nu}\,S'_{\nu}\,{b'}^\dag_{t+\tau_\nu}+ {\rm H.c.}\label{Vt-rl}
\end{equation}
with $\gamma=g_{k_0}^2/ v$ and $\gamma'={g}_{-k_0}^2/ v$ (which we allow to be generally different). Here, $b_t$ are right-going time modes fulfilling $[b_t,b_{t'}^\dag]=\delta(t-t')$, $[b_t,b_{t'}]=[b^\dag_t,b_{t'}^\dag]=0$. Likewise, $b'_t$ define left-going modes with analogous commutation rules. 
Before proceeding further, recalling \eq\eqref{Snu}, it is convenient to define the collective atomic operators 
\begin{eqnarray}
S&=&\sum_\nu S_{\nu}=\sum_{j}\left(\sum_{\ell}e^{-i\varphi_{j\ell}}\right)\sigma_{j}\,,\label{coll1}\\
S'&=&\sum_\nu S'_{\nu}=\sum_{j}\left(\sum_{\ell}e^{i\varphi_{j\ell}}\right)\sigma_{j}\,.\label{coll2}
\end{eqnarray}

The regime of negligible time delays is defined by $\tau_{\cal N}-\tau_1\ll \gamma^{-1},{\gamma'}^{-1}$, allowing to coarse grain the dynamics over a characteristic time scale $\Delta t$ such that 
\begin{equation}
\tau_{\cal N}-\tau_1\ll \Delta t\ll\gamma^{-1},{\gamma'}^{-1}\label{hie}\,.
\end{equation}
In this regime, it can be shown \cite{CP} that \eqref{Vn} and \eqref{Hn} reduce to 
	\begin{eqnarray}
\overline V_{n}&=&\tfrac{1}{\sqrt{\Delta t}}\,\left(\sqrt{\gamma}\,S  \,b_n^\dag+\sqrt{\gamma'}\,S' \, {b'}_n^\dag+{\rm H.c.}\right),\,\,\,\,\,\label{Vn-rl}	\\
 {\cal H}_n&=&\tfrac{i}{2}\sum_{\nu>\nu'}\left( \gamma\, S_{\nu'}^\dag {S}_{\nu}+ \gamma' {S'}_{\nu}^\dag {S'}_{\nu'}-{\rm H.c.}\right),\label{Hvac-rl}
\end{eqnarray} 
where
\begin{equation}
b_n=\tfrac{1}{\sqrt{\Delta t}}\int_{t_{n-1}}^{t_n}\!dt\, b_t\,,\,\,\,b'_n=\tfrac{1}{\sqrt{\Delta t}}\int_{t_{n-1}}^{t_n}\!dt\, b'_t\,.\label{bn}
\end{equation}
\eq\eqref{bn} define a discrete set of ladder operators of the environment fulfilling bosonic commutation rules $[b_n,b_{n'}^\dag]=\delta_{n,n'}$, $[b_n,b_{n'}]=[b^\dag_n,b_{n'}^\dag]=0$ (and likewise for $b_n'$), as is easily checked using the commutation rules of time-mode operators $b_t$ and $b'_t$.

Based on \eqref{Vn-rl}, the DF condition [\cf\eq\eqref{cond}] for a bidirectional waveguide in terms of collective atomic operators \eqref{coll1} and \eqref{coll2} simply reads 
\begin{equation}
S=S'=0\,.\label{S-Sp}
\end{equation}
This is equivalent to \eqref{cond-g2} [$S=0\Leftrightarrow S'=0$ since if \eq\eqref{cond-g2} is matched so is the analogous equation for $\varphi_{j\ell}\rightarrow -\varphi_{j\ell}$].

\subsection{Effective Hamiltonian}\label{sec-eh}

When $\overline V_n=0$, atoms will evolve unitarily with effective Hamiltonian [recall \eq\eqref{Heff}] ${\cal H}_n=H_{\rm eff}\otimes \openone$ (${\cal H}_n$ acts trivially on the field). Henceforth, we will omit the identity operator.

The effective Hamiltonian $H_{\rm eff}={\cal H}_n$ can be written more explicitly as [\cf\eqs\eqref{Snu} and \eqref{Hvac-rl}]
\begin{align}
H_{\rm eff}\,=\,&\tfrac{i}{2}\sum_{\nu>\nu'}\left( \gamma\,  e^{i (\varphi_{j'\ell'}-\varphi_{j\ell})} \sigma_{j'}^\dag\sigma_{j}\right.\nonumber\\
&\left.\,\,\,\,\,\,\,\,\,\,\,\,\,\,+\gamma'\,  e^{i (\varphi_{j'\ell'}-\varphi_{j\ell})} \sigma_{j}^\dag\sigma_{j'}-{\rm H.c.}\right)\,,
\end{align}
where ($j$, $\ell$) are understood as the pair of indexes corresponding to $\nu$ [and likewise ($j'$, $\ell'$) with respect to $\nu'$]. This
in turn can be expressed in the compact form
\begin{align}
H_{\rm eff}=\sum_{jj'}J_{jj'}\,\sigma_{j}^\dag\sigma_{j'}+{\rm H.c.}\,\label{Heff2}
\end{align}
with 
\begin{align}
J_{jj'}&=\sum_{\nu_{j'\ell'}>\nu_{j\ell}}\left[\tfrac{\gamma+\gamma'}{2}\sin(\varphi_{j'\ell'}-\varphi_{j\ell})\right.\nonumber\\
&\,\,\,\,\,\,\,\,\,\,\,\,\,\,\,\,\,\,\,\,\,\,\,\,\,\,\,\,\,\,\left.+i\,\tfrac{\gamma-\gamma'}{2}\cos{(\varphi_{j'\ell'}}-\varphi_{j\ell})\right]\,,\label{jjj}
\end{align}
and where $\nu_{j\ell}$ is the (previously introduced) discrete map returning the coupling point index for each pair $(j,\ell)$. Note that for isotropic coupling ($\gamma=\gamma'$), each $J_{jj'}$ (for given $j$ and $j'$) reduces to a sum of sine functions, where the argument of each sine is the phase shift associated to a pair of coupling points (one of atom $j$ one of $j'$).
Alternatively, $J_{jj'}$ can be expressed by separating the right- and left-going contributions as
\begin{equation}
J_{jj'}=\gamma K_{jj'}+\gamma'  K^*_{jj'}\label{jjj2}
\end{equation}
with
\begin{equation}
K_{jj'}=\tfrac{1}{2}\!\sum_{\nu_{j'\ell'}>\nu_{j\ell}}e^{i(\varphi_{j\ell}-\varphi_{j'\ell'}+\frac{\pi}{2})}\,.
\end{equation}

\begin{figure}
	\includegraphics[width=0.48\textwidth]{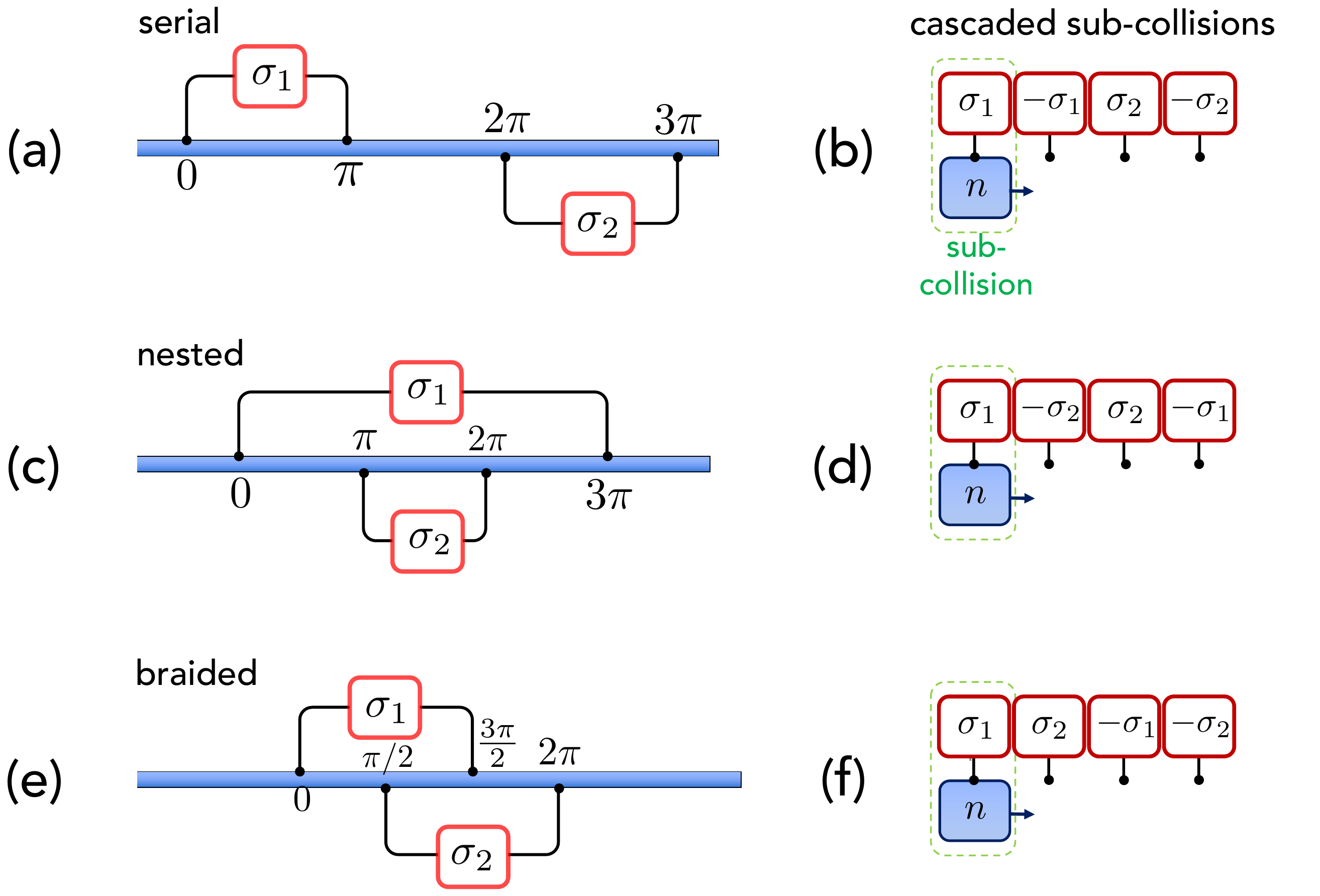}
	\caption{Coupling point topologies of two giant atoms each with two coupling points (in a unidirectional waveguide) and corresponding sequences of cascaded subcollisions. (a) and (b): serial topology and corresponding sequence of $S_\nu$ [\cf\eq\eqref{Snu}]: $S_1=\sigma_1$, $S_2=-\sigma_1$, $S_3=\sigma_2$, $S_4=-\sigma_2$. (c) and (d): nested topology. (e) and (f): braided topology. The phase $k_0 x_\nu$ of each coupling point is shown in (a), (c), (e). Each subcollision is described by the unitary $e^{-i V_{n\nu} \Delta t}$ with $V_{n\nu}$ given by  \eq\eqref{Vnnu}. In (f), we absorbed a phase factor $e^{-i \pi/2}$ in the definition of $\sigma_2$.}\label{topo}
\end{figure}

\begin{figure}
	\includegraphics[width=0.46\textwidth]{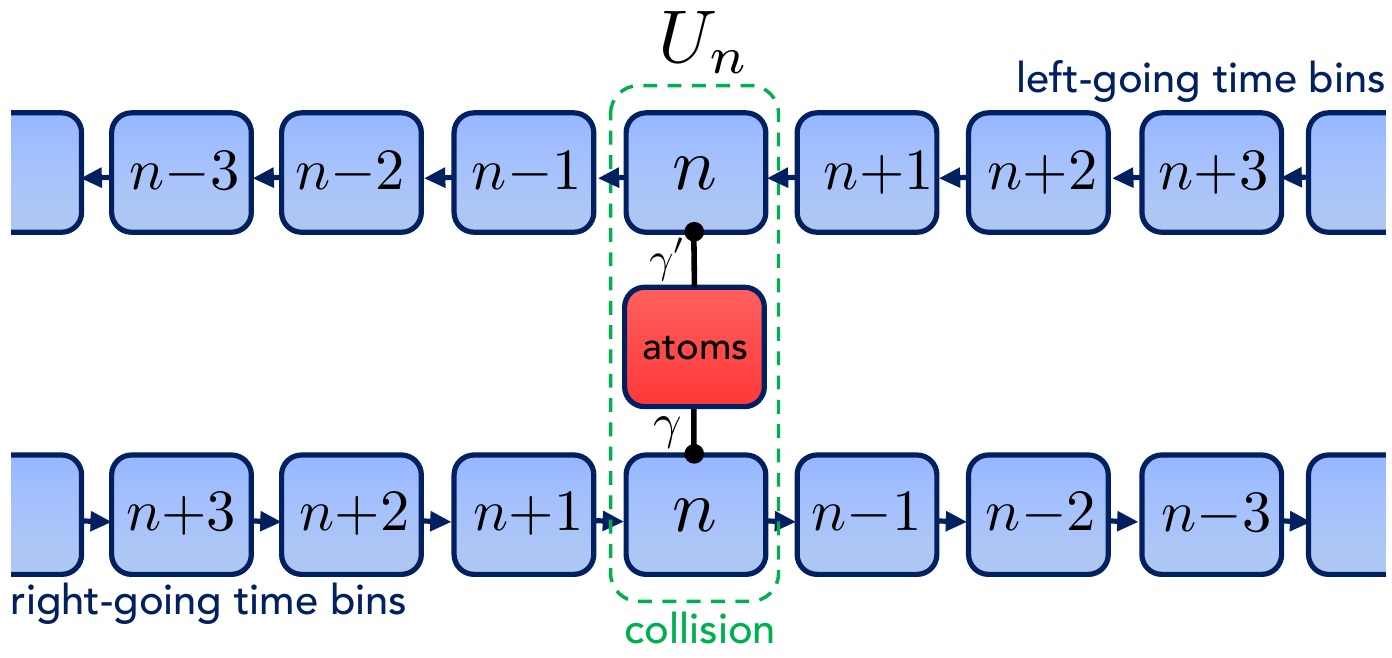}
	\caption{Collisional picture of the joint dynamics of giant atoms and field. The field is decomposed into non-interacting time bins, each generally corresponding to the pair of harmonic oscillators $(b_n,b'_n)$. At step $n$ (time interval $[t_{n-1},t_n]$) the $n$th time bin collides with all the atoms according to coupling Hamiltonian $\overline V_n$ [\cf\eq\eqref{Vn-rl}]. The atoms at the same time are subject to an internal coherent dynamics described by Hamiltonian ${\cal H}_n\equiv H_{\rm eff}$ [\cf\eq\eqref{Hvac-rl}]. Both ${\cal H}_n$ and $\overline V_n$ are encompassed in the pairwise collision unitary $U_n$ [\cf\eqs\eqref{Ut} and \eqref{Un-app}]. In the unidirectional case $\gamma'=0$, only right-going time bins $b_n$ are to be considered.}\label{macro}
\end{figure}

The issue is now raised as to whether or not $H_{\rm eff}\neq 0$ when decoherence is inhibited [condition \eqref{cond-g2}]. It turns out that there generally exist patterns of coupling points such that $H_{\rm eff}= 0$ and patterns for which $H_{\rm eff}\neq 0$, where the former yield a trivial dynamics (the system just does not evolve) and are thus unwanted.
The best instance for illustrating this is a pair of giant atoms 1 and 2, such that ${\cal N}_1={\cal N}_2=2$, with equally-spaced coupling points 
\begin{equation}
k_0 x_\nu=(\nu-1) \varphi\,\,\,\,{\rm with}\,\,\,\, \nu=1,2,3,4\,,\label{setting}
\end{equation}
the field being unidirectional ($\gamma'=0)$. Three different types of patterns are then possible: serial, nested and braided [see \fig\ref{topo}(a), (c), (e)]. For the serial and nested topology, we choose $\varphi=\pi$ while in the braided case we take $\varphi=\pi/2$. Each of these settings ensures that there is $(2n{+}1)\pi$-phase-shift between the two coupling points of each giant atom, thus matching the DF condition [recall \eq\eqref{cond-g2}]. Using \eqref{Heff2}, in the serial and nested topologies we get $H_{\rm eff}=0$, while the braided yields \cite{KockumPRL2018}
\begin{equation}
H_{\rm eff}=\gamma\, (\sigma_1 \sigma_2^\dag +\sigma_1^\dag \sigma_2)\,,\label{Heff-bra}
\end{equation}
(we absorbed a phase factor $e^{-i \pi/2}$ in the definition of $\sigma_2$).
Analogous conclusions hold for isotropic coupling ($\gamma'=\gamma=\Gamma/2$), in which case \eqref{Heff-bra} is generalized by replacing $\gamma$ with $\Gamma$.
\\
\\
\indent Before concluding this section, we recall that, as discussed in \rref\cite{CP}, based on \eqs\eqref{Ut}, \eqref{Un-app}, \eqref{Vn-rl} and \eqref{Hvac-rl} one can effectively see the joint dynamics as a collision model \cite{ciccarelloCollision2017,brunSimple2002,altamiranoUnitarity2017,lorenzoComposite2017,grossQubit2018} (see \fig\ref{macro}). According to this (we first consider the unidirectional case $\gamma'=0$), the field is decomposed into a discrete stream of right-going time bins, each being a bosonic mode with ladder operator $b_n$ defined by \eq\eqref{bn}. During the time interval $[t_{n-1},t_n]$, the $n$th time bin undergoes a collision with all the atoms at once, which is described by unitary $U_n$ [\cf\eq\eqref{Un-app}]. The atoms at the same time are effectively subject to a mutual coherent interaction described by ${\cal H}_n\equiv H_{\rm eff}$ (encompassed in the collision unitary $U_n$). The extension to the bidirectional case is natural: there is now an additional stream of left-going time bins (each with ladder operator $b'_n$). One can equivalently think of two-mode time bins $(b_n,b'_n)$ such that in each collision the atoms collide with both time-bin subsystems $b_n$ and $b'_n$ (see \fig\ref{macro}).

\section{Mapping the giant atoms dynamics into a cascaded collision model}\label{CM-def}

While, as pointed out in the previous section, the collision $U_n$ formally describes a simultaneous collision with all the atoms, we show next that it can be effectively decomposed as a {\it cascade} of sub-collisions each involving only {\it one} coupling point. Cascaded collision models (for normal atoms) were introduced in \rrefs\cite{giovannettiMaster2012a,giovannettiMaster2012} (see also \rref\cite{LorenzoFlux}). Here, a unidirectional waveguide is considered, the extension to the bidirectional case being postponed to Sec.~\ref{sec-bid}.

\begin{figure}
	\includegraphics[height=0.45\textwidth]{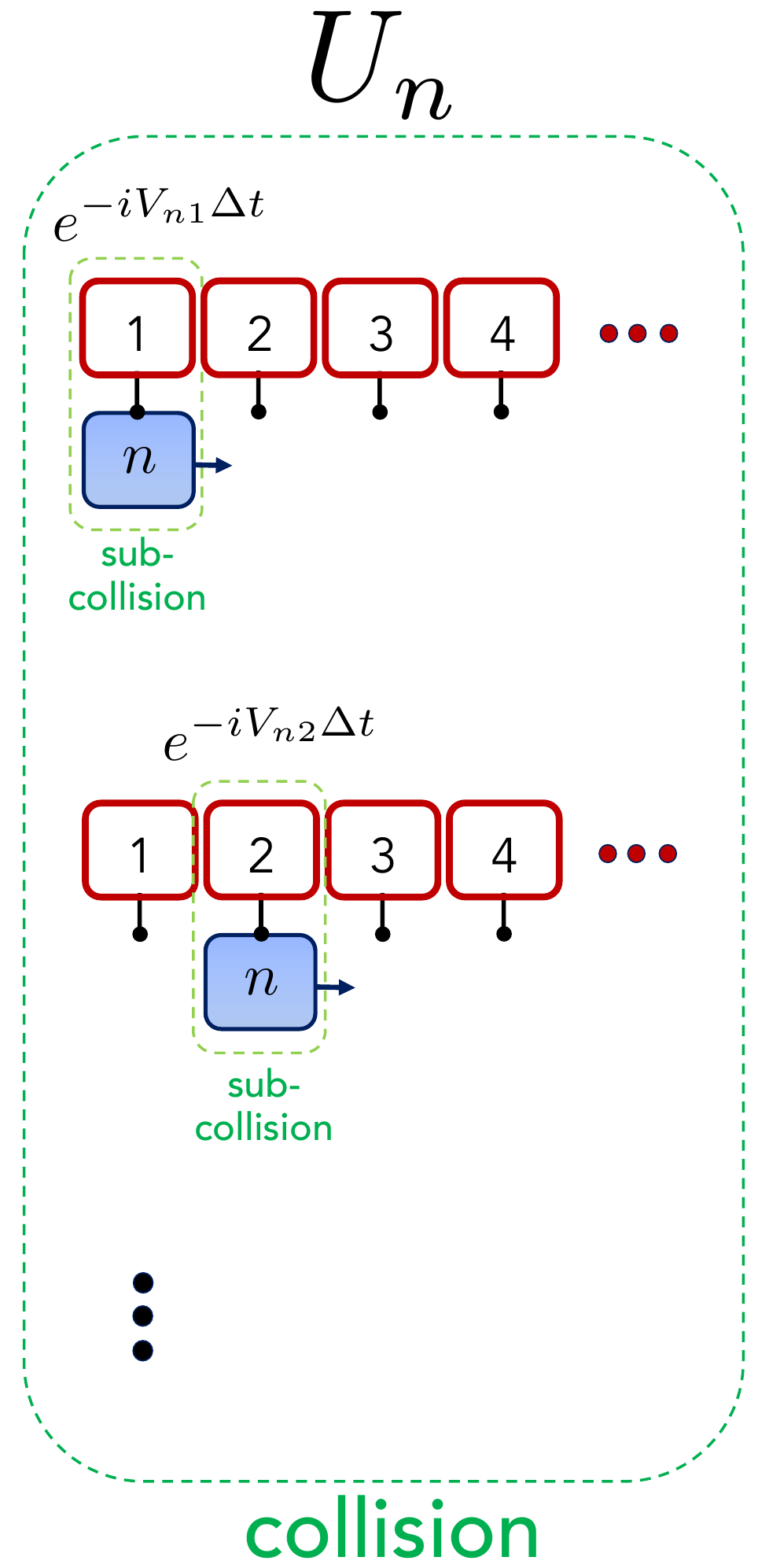}
	\caption{Under the DF condition \eqref{cond-g2}, each collision in \fig\ref{macro} (we address here the case $\gamma'=0$) can be effectively decomposed into ${\cal N}$ cascaded sub-collisions according to \eq\eqref{deco}. Each sub-collision is between the same time bin $n$ and a different coupling point corresponding to coupling Hamiltonian \eqref{Vnnu} (time grows from top to bottom).}\label{sub}
\end{figure}

For each coupling point $\nu$, let us define the interaction Hamiltonian 
\begin{equation}
 V_{n\nu}= \sqrt{\tfrac{\gamma}{\Delta t}}  \, \left( {S}_{\nu}\, b_{n}^\dag+{\rm H.c.}\right)\,\label{Vnnu}
\end{equation}
coupling the $n$th time bin to atom $j$ with phase $\varphi_{j\ell}$ [\cf\eq\eqref{Snu}], where $(j,\ell)$ is the pair correspond to coupling point $\nu$ (in the remainder we introduce a convenient terminology and say that the time bin ``interacts with the coupling point"). 
Using \eqs\eqref{coll1}, \eqref{coll2}, \eqref{Vn-rl} and \eqref{Vnnu}, it is easily immediately checked that the average interaction Hamiltonian is just the sum of the $V_{n\nu}$'s
\begin{equation}
\overline V_n=\sum_\nu V_{n\nu} \label{Vmed}\,.
\end{equation}
More importantly, as shown in the remainder, it turns out that, when the DF condition \eqref{cond-g2} is matched, the unitary collision $U_n$ can be decomposed as
\begin{equation}
{U}_n=e^{-i  V_{n{\cal N}} \Delta t}\cdots \,e^{-i  V_{n1} \Delta t}\label{deco}\,.
\end{equation}
Thereby, one can think of each collision (see \fig \ref{sub}) as the result of ${\cal N}$ cascaded sub-collisions in each of which the time bin ``collides" with one of the coupling points according to unitary $e^{-i  V_{n\nu} \Delta t}$ with $V_{n\nu}$ given by \eqref{Vnnu}. Of course, this in particular entails that the same time bin collides with a given atom as many times as the number of respective coupling points ${\cal N}_j$. Yet, the sub-collisions with the same atom occur with different coupling Hamiltonians and are generally non-consecutive (i.e., between two sub-collisions with the same atom $j$ there may be sub-collisions with atoms $j'\neq j$), which is key to the occurrence of a non-trivial DF Hamiltonian as we will see shortly.

To prove \eqref{deco}, we expand to second order each sub-collision unitary on the right hand side as $e^{-i  V_{n\nu} \Delta t}\simeq \openone-i V_{n\nu} \Delta t-\tfrac{1}{2}V_{n\nu}^2 \Delta t^2$. This yields (to leading order) 
\begin{equation}
\prod_{\nu=1}^{{\cal N}}\,e^{-i  V_{n\nu} \Delta t}\simeq  \openone  -i\, (\overline V_{n}+\tilde{\cal H}_n)\, \Delta t-\tfrac{1}{2}\,\overline V_{n}^2\,\Delta t^2\label{2ndord}
\end{equation}
with the order in the product understood as in \eqref{deco} and
\begin{equation}
\tilde{\cal H}_n=i\,\tfrac{\Delta t}{2}\! \sum_{\nu>\nu'}\left[V_{n\nu'},V_{n\nu}\right]\,,\label{H2bis}
\end{equation}
where we used \eqref{Vmed}. Using \eqref{cond-g2}, it is easily shown that $\tilde{\cal H}_n={\cal H}_n\equiv H_{\rm eff}$ (see Appendix \ref{AppA}). Upon comparison with \eqref{Un-app}, we thus conclude that \eqref{deco} holds true.

The decomposition in terms of cascaded sub-collisions in particular highlights the physical origin of effective Hamiltonian \eqref{Hvac-rl}: if, instead of being sequential, the sub-collisions occurred simultaneously (corresponding to perfectly co-located coupling points) then the overall collision unitary would be $e^{-i (\sum_\nu\! V_{n\nu})\Delta t }\equiv e^{-i \overline V_{n}\Delta t }$, the corresponding second-order expansion being just \eqref{2ndord} {\it without} term $\tilde{\cal H}_n\equiv H_{\rm eff}$. Thus the effective Hamiltonian arises precisely because the time bin collides with the coupling points in a cascaded fashion. This is in fact the same mechanism underpinning emergence of effective Hamiltonians in chiral quantum optics with normal atoms \cite{LodahlReviewNature17}, the difference yet being that decoherence cannot be suppressed in the latter case (because $\overline V_{n}$ cannot vanish with normal atoms).

\section{Mechanism behind emergence of  non-trivial $H_{\rm eff}$}\label{sec-mec}

Occurrence of non-trivial (i.e., non-zero) DF Hamiltonians is simply interpreted in the cascaded-collision-model picture. 

As in Sec.~\ref{sec-eh}, throughout this and the next section we consider a {\it unidirectional} waveguide and giant atoms with {\it two coupling points} each, which captures most of the essential physics.
\subsection{Single Giant Atom} 
Let us consider first a single giant atom
and set $x_1=\tau_1=0$, $\varphi=k_0 x_2$. The DF condition $\overline{V}_n=0$ then simply reads $\varphi=(2n{+}1)\pi$ with $n$ an integer number. Hence, $S_1=-S_2=\sigma_1$ and [\cf\eq\eqref{Vnnu}]
\begin{equation}
V_{n1}=\sqrt{\tfrac{\gamma}{\Delta t}}\,( \sigma_{1}\, b_{n}^\dag+{\rm H.c.})\,,\,\,V_{n2}=\sqrt{\tfrac{\gamma}{\Delta t}}\,( -\sigma_{1}\, b_{n}^\dag+{\rm H.c.})\,.
\end{equation}
Thus $V_{n1}=-V_{n2}$ and [see \eq\eqref{deco}]
\begin{equation}
U_n=e^{-i V_{n2}\Delta t}\,e^{-i V_{n1}\Delta t}=\openone\,,\label{id}
\end{equation}
meaning that the collision has no effect overall. This, in particular, necessarily entails $H_{\rm eff}=0$ [recall \eq\eqref{Un-app}]. In other words, the two sub-collisions are the {\it time-reversed} of one another (so that the net effect is null). To sum up, in order to ensure the DF condition $\overline{V}_n=0$ for a single giant atom, one must adjust the phase shift so that $V_{n2}=-V_{n1}$. This yet brings about that one sub-collision is just the other one time-reversed, trivially yielding $U_n=0$ hence $H_{\rm eff}=0$.

\subsection{Two Atoms}
When it comes to a pair of giant atoms, instead, conditions $\overline{V}_n=0$ and $U_n\neq 0$ can be matched simultaneously. To see this, we reconsider uniformly-spaced atoms as in \eq\eqref{setting} and always set $\varphi$ so as to ensure a $(2n{+}1)\pi$-phase-shift between the pair of coupling points of each atom, hence $\overline{V}_n=0$ (similarly to the single-atom instance just discussed). 

For convenience, we define the coupling Hamiltonians
\begin{equation}
\mathcal V_{j}=\sqrt{\tfrac{\gamma}{\Delta t}}\left( \sigma_{j}\, b_{n}^\dag+{\rm H.c.}\right)\label{calVj}
\end{equation}
with $j=1,2$ (the dependence on $n$ is left implicit). No phase factor appears in this definition.

Consider first the serial scheme in \fig \ref{topo}(a), in which case we set $\varphi=\pi$. Then [see \fig\ref{topo}(b)], 
\begin{equation}
V_{n1}={\mathcal V}_1\,,\,\,V_{n2}=-{\mathcal V}_1\,,\,\,V_{n3}={\mathcal V}_2\,,\,\,V_{n4}=-{\mathcal V}_2\,.
\end{equation}
This results in the collision unitary [\cf\eqref{deco}]
\begin{equation}
U_n=e^{i {\cal V}_2 \Delta t}e^{-i {\cal V}_2 \Delta t}e^{i {\cal V}_1 \Delta t}e^{-i {\cal V}_1 \Delta t}=\openone\,,
\end{equation}
that is a trivial dynamics such that $H_{\rm eff}=0$. This case is in fact an extension the single giant atom considered above.

For the nested case in \fig\ref{topo}(c), we set $\varphi=\pi$. Then [see \fig\ref{topo}(d)], 
\begin{equation}
V_{n1}={\mathcal V}_1\,,\,\,V_{n2}=-{\mathcal V}_2\,,\,\,V_{n3}={\mathcal V}_2\,,\,\,V_{n4}=-{\mathcal V}_1\,,
\end{equation}
Thus the second pair of sub-collisions is the first pair time-reversed
\begin{equation}
U_n=e^{i {\cal V}_1 \Delta t}e^{-i {\cal V}_2 \Delta t}e^{i {\cal V}_2 \Delta t}e^{-i {\cal V}_1 \Delta t}=\openone\,,
\end{equation}
ensuing again a trivial dynamics and $H_{\rm eff}=0$. Equivalently, the pair of central sub-collisions, both involving atom 2, are the time-reversed of one another. Thus atom 2 simply disappears from $U_n$, which reduces to $U_n=e^{i {\cal V}_1 \Delta t}e^{-i {\cal V}_1 \Delta t}=\openone$.

For the braided arrangement of \fig\ref{topo}(e), we set $\varphi=\pi/2$. Then [see \fig\ref{topo}(f)], 
\begin{equation}
V_{n1}={\mathcal V}_1\,,\,\,V_{n2}={\mathcal V}_2\,,\,\,V_{n3}=-{\mathcal V}_1\,,\,\,V_{n4}=-{\mathcal V}_2
\end{equation}
[with $\mathcal V_{2}$ now defined by \eqref{calVj} for $j=2$ under the replacement $\sigma_{2}\rightarrow -i \sigma_{2}$]. The collision unitary is given by
\begin{equation}
U_n=e^{i {\cal V}_2 \Delta t}e^{i {\cal V}_1 \Delta t}e^{-i {\cal V}_2 \Delta t}e^{-i {\cal V}_1 \Delta t}=e^{-i \gamma (\sigma_1 \sigma_2^\dag +\sigma_1^\dag \sigma_2) \Delta t}\neq \openone\,.\label{Un-bra}
\end{equation}
Therefore, $\overline V_n=0$ is fulfilled but now $H_{\rm eff}\neq 0$.

The above shows that, while being irrelevant for realizing the DF condition $\overline V_n=0$, the coupling points topology is crucial in order to have a non-vanishing effective Hamiltonian. In terms of propagators [\cf\eqs\eqref{Ut} and \eqref{Un-app}], this is ultimately due to the fact that the second-order term ${\cal H}_n$ is affected by the time-ordering operator, while ${\overline V_n}$ and (of course) ${\overline V_n}\!^2$ are fully insensitive to it.

\subsection{Many atoms}\label{many}

The above arguments are naturally extended to more than two giant atoms. 
Again, the DF condition \eqref{cond-g2} is matched when for each atom the phase shift between its coupling points is a multiple integer of $\pi$, that is $\varphi_{j,2}-\varphi_{j,1}=(2n_j+1)\pi$ for some integer $n_j$ and for any $j$. 
Based on the above discussion, we can state that if there exists an atom $j$ such that no coupling point of other atoms lies between its coupling points, i.e., $x_{j',\ell'}\not\in [x_{j,1},x_{j,2}]$ for all $j'\neq j$ and $\ell'=1,2$, then $H_{\rm eff}$ simply does not contain $\sigma_{j}$ and $\sigma_{j}^\dag$ (i.e., atom $j$ is fully decoupled from the field and other atoms). This is because the pair of sub-collision unitaries corresponding to coupling points $x_{j,1}$ and $x_{j,2}$ are the time-reversed of one another, hence atom $j$ is fully decoupled from the field and all other atoms. A braided configuration is thus generally defined \cite{KockumPRL2018} as one such that no atom exhibits the above phenomenon.



\section{Equivalent scheme using a mediator} \label{sec-qubit}

Performing quantum information processing tasks \cite{Nielsen00} on two {\it remote} systems, say $1$ and $2$, is a longstanding problem. The challenge is due to the impossibility of acting  jointly on the systems (owing to their distance). A standard strategy to get around this is employing a mobile mediator: a third quantum system $M$ which can shuttle between 1 and 2 so as to mediate an indirect coupling, which, e.g., can be exploited for generating entanglement. 
\subsection{Two-qubit gate}
The dynamics realizing the DF Hamiltonian for two giant atoms in a braided configuration [see \fig\ref{topo}(e)] in fact implements a two-qubit gate via a shuttling mediator. To see this, for simplicity and in view of the equivalent quantum circuit to be discussed later, we consider the field initially in the vacuum state, in which case each time bin behaves as an effective qubit $M$ \cite{note1}.
Then ${\cal V}_j$ [\cf\eq\eqref{calVj}] reads
\begin{equation}
\mathcal V_{j}=\sqrt{\tfrac{\gamma}{\Delta t}}\left( \sigma_{j}\, \sigma_{M+}+\sigma_{j}^\dag\, \sigma_{M-}\right)\,.\label{calVj-2}
\end{equation}
with $\sigma_{M-}=\sigma_{M+}^\dag$ usual spin-1/2 ladder operators of $M$. Note that, for convenience, we gauged away a phase factor $e^{-i \pi/2}$ [due to $\varphi=\pi/2$, \cf\eq\eqref{setting}] by absorbing it in the definition of $\sigma_2$.
\begin{figure}
	\includegraphics[width=0.43\textwidth]{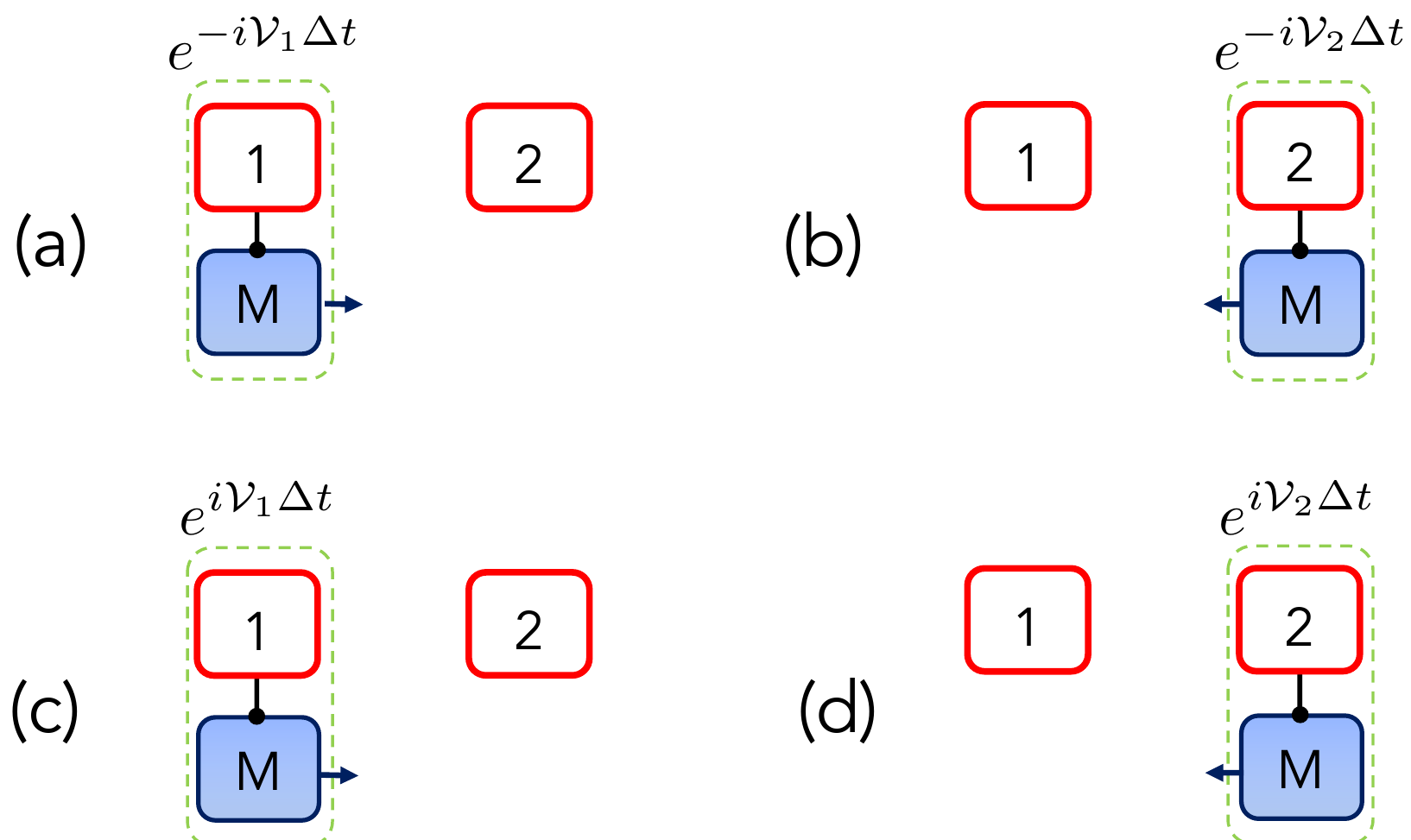}
	\caption{The four subcollisions in the braided configuration can be seen as a mediator (embodied by the field time bin) shuttling between atoms 1 and 2. This represents a collision model where in each collision the ancilla $M$ (time-bin) interacts twice with each subsystem $j=1,2$ the first time with coupling Hamiltonian ${\cal V}_j$ [\cf\eq\eqref{calVj-2}] and the second time with $-{\cal V}_j$. The net result is a coherent 1-2 interaction with $M$ eventually returning to its initial state. By applying phase kicks on $M$, one can make both subcollisions with $j$ occur with the same coupling ${\cal V}_j$ [see \fig\eqref{circuit}]}.\label{shuttle}
\end{figure}

Unitary \eqref{Un-bra} (braided configuration) can be interpreted as follows (see \fig\ref{shuttle}). $M$ is first put close to 1 and the $M$-1 two-qubit gate $e^{-i {\cal V}_1 \Delta t}$ applied on them [\fig\ref{shuttle}(a)], then $M$ is moved close to 2 and $M$-2 gate $e^{-i {\cal V}_2 \Delta t}$ applied (b), next $M$ goes back to 1 and gate $e^{i {\cal V}_1 \Delta t}$ is applied (c). Finally, $M$ lies close to 2 again and gate $e^{i {\cal V}_2 \Delta t}$ is applied (d). 

Consider now the more realistic case that only one two-qubit gate per atom can be implemented, say $e^{-i {\cal V}_1 \Delta t}$ and $e^{-i {\cal V}_2 \Delta t}$: we ask whether the other two can be obtained from these by adding extra single-qubit (local) gates. Noting that the local unitary transformation defined by $U_M=\sigma_{Mz}$ transforms the $M$'s ladder operators as $\sigma_{M\pm}\rightarrow -\sigma_{M\pm}$, we have 
\begin{equation}
\sigma_{Mz} \, {\cal V}_j \,\sigma_{Mz}=-{\cal V}_j\,\Rightarrow\,e^{i {\cal V}_j \Delta t}=\sigma_{Mz} \,e^{-i {\cal V}_j \Delta t}\,\sigma_{Mz}\,
\end{equation}
with $j=1,2$. Plugging  the last decomposition of $e^{i {\cal V}_j \Delta t}$ in \eqref{Un-bra} thus yields
\begin{equation}
U_n=\sigma_{Mz} \,e^{-i {\cal V}_2 \Delta t} \,e^{-i {\cal V}_1 \Delta t}\,\sigma_{Mz}\, e^{-i {\cal V}_2 \Delta t}e^{-i {\cal V}_1 \Delta t}\,,
\end{equation}
where we used $\sigma_{Mz}^2=1$. Recalling that in the braided configuration, $U_n=e^{-i H_{\rm eff} \Delta t}$ with $H_{\rm eff}$ given by \eqref{Heff-bra}, we conclude
\begin{equation}
e^{-i \gamma (\sigma_1 \sigma_2^\dag +{\rm H.c.}) \Delta t}=\sigma_{Mz} \,e^{-i {\cal V}_2 \Delta t} \,e^{-i {\cal V}_1 \Delta t}\,\sigma_{Mz}\, e^{-i {\cal V}_2 \Delta t}e^{-i {\cal V}_1 \Delta t}\,.\label{deco2}
\end{equation}
The collision unitary in the braided configuration -- hence the entire dynamics in fact -- can thus be seen as alternate subcollisions (of the same type) of $M$ with 1 and 2 where a local $\pi$-phase gate is applied on $M$ at the end of each cycle. This effectively implements a DF interaction between 1 and 2 only.
Note that replacing $\sigma_{Mz}={\rm diag}(1,e^{i \pi})$ (written in matrix form) with another phase gate $U_{M}={\rm diag}(1,e^{i \varphi})$ with $\varphi\neq (2n+1)\pi$ generally gives rise to an overall unitary $U_n$ which does {\it not} act trivially on $M$, thus introducing decoherence. 

The above is somewhat reminiscent of dynamical decoupling schemes \cite{ViolaPRL1999}, where suitable local pulses are repeatedly applied in order to effectively decouple the system from the environment thus suppressing decoherence. Note however that in our case the environment (embodied by $M$, that is the field in fact) has an active role since it allows 1 and 2 to effectively crosstalk.

\subsection{Quantum circuit}

To express \eqref{deco2} in the language of quantum circuits \cite{Nielsen00}, we recall the definition of an ${\rm XY}$ gate, also known as piSWAP or parametric i-SWAP,
\begin{equation}
{\rm XY}(\delta) =
\begin{bmatrix}
1 & 0 & 0 & 0\\
0 & \cos{\pi\delta} & -i\sin{\pi\delta} & 0\\
0 & -i\sin{\pi\delta} & \cos{\pi\delta} & 0\\
0 & 0 & 0 & 1\\
\end{bmatrix} \,,
\end{equation}
which coincides with each $M$-$j$ unitary $e^{-i {\cal V}_j\Delta t}$ for $\delta =\tfrac{1}{\pi}\sqrt{\gamma \Delta t}$. 
\begin{figure}
	\includegraphics[width=0.45\textwidth]{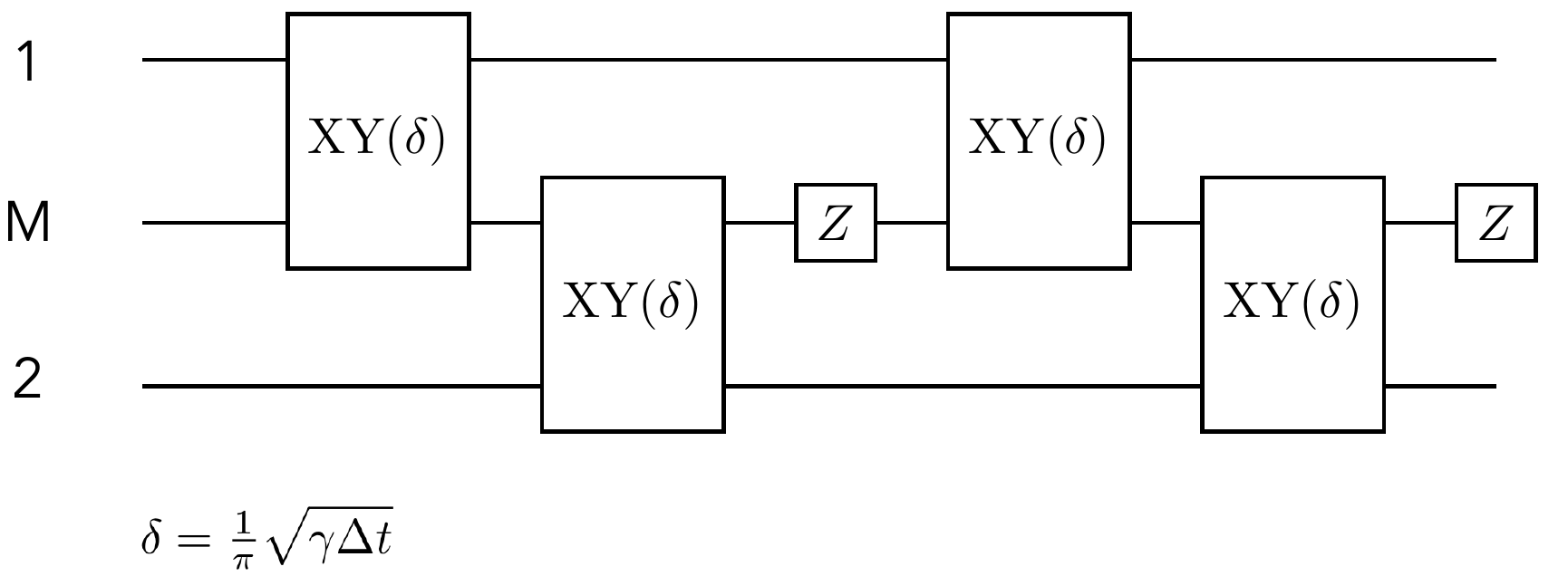}
	\caption{Equivalent quantum circuit of the unitary collision in the braided scheme [\cf\eqs\eqref{Un-bra} and \eqref{deco2}]. A (maximally entangling) i-SWAP two-qubit gate on 1 and 2 can be effectively obtained by iterating this quantum circuit.}\label{circuit}
\end{figure}
Thereby we get the equivalent quantum circuit of the collision unitary \eqref{Un-bra} displayed in \fig\ref{circuit} (where $Z=\sigma_z$). 

Thus, remarkably, the waveguide setup with giant atoms can be seen as implementing iterated applications of the elementary quantum circuit in \fig\ref{circuit}. A canonical (maximally entangling) i-SWAP two-qubit gate $U_n=e^{-i (\sigma_1 \sigma_2^\dag +{\rm H.c.})}$ is obtained after $N=[(\gamma \Delta t)^{-1}]$ iterations [recall \eq\eqref{hie}].

{\section{Bidirectional chiral waveguide}\label{sec-bid}

The considerations in Sec.~\ref{sec-mec} naturally extend to a bidirectional waveguide. In Sec.~\ref{many}, we saw that, in the unidirectional case, if an atom $j$ is  such that $x_{j',\ell'}\not\in [x_{j,1},x_{j,2}]$ for any $j'\neq j$ and $\ell'=1,2$ (configuration {\it not} braided) then $\sigma_{j}$ and $\sigma_{j}^\dag$ do not appear in $H_{\rm eff}$ under the DF condition. The same holds for a bidirectional waveguide since, according to \eqs\eqref{Heff2} and \eqref{jjj2}, if $J_{jj'}$ vanishes in the unidirectional case $\gamma'=0$ then so does in the bidirectional one (when $\gamma'\neq 0$).

For completeness, it is however worth showing that one can reach the same conclusion even through a purely collisional argument. To this aim, we note that, even for a bidirectional waveguide, under the DF condition \eqref{cond-g2} the collision unitary [\cf\eqs\eqref{Un-app}, \eqref{Vn-rl}, \eqref{Hvac-rl}] can be decomposed into cascaded sub-collisions [\cf\eq\eqref{deco}] as 
\begin{equation}
{U}_n=e^{-i  (V_{n{\cal N}}+V'_{n{1}}) \Delta t}\cdots \,e^{-i  (V_{n1}+V'_{n{\cal N}})\Delta t}\label{deco3}\,.
\end{equation}
with [\cf\eq\eqref{Vnnu}]
\begin{equation}
V'_{n\nu}= \sqrt{\tfrac{\gamma'}{\Delta t}}  \, \left( {S'}\!_{\nu}\, {b'}_{n}^\dag+{\rm H.c.}\right)\,\label{Vnnu-2}
\end{equation}
[recall \eqs\eqref{Snu} and \eqref{bn}].
This is because [\cf\eq\eqref{Un-app}] $\overline V_n=\sum_\nu (V_{n\nu}{+}V'_{n\nu})$ and, as shown in Appendix \ref{AppA}, $\tilde{\cal H}_n={{\cal H}_{n}}\equiv H_{\rm eff}$. Here, $\tilde{\cal H}_n$ [\cf\eq\eqref{H2bis}] is now generally defined as
\begin{align}
\tilde{\cal H}_n=\! i \tfrac{\Delta t}{2}  \sum_{\nu > \nu'}\left[V_{n,\nu'}{+}V'_{n,\,{\cal N}+1-\nu'},V_{n,\nu}{+}V'_{n,\,{\cal N}+1-\nu}\right] \,.\label{comm-2way}
\end{align}
Note that, as sketched in \fig\ref{sub2}, in the first sub-collision, the right-going time bin interacts with coupling point $\nu=1$ and the left-going time bin with coupling point $\nu={\cal N}$. Then, in the second sub-collision, the right-(left-)going time bin interacts with coupling point $\nu=2$ ($\nu={\cal N}{-}2$) and so on.
\begin{figure}
	\includegraphics[height=0.47\textwidth]{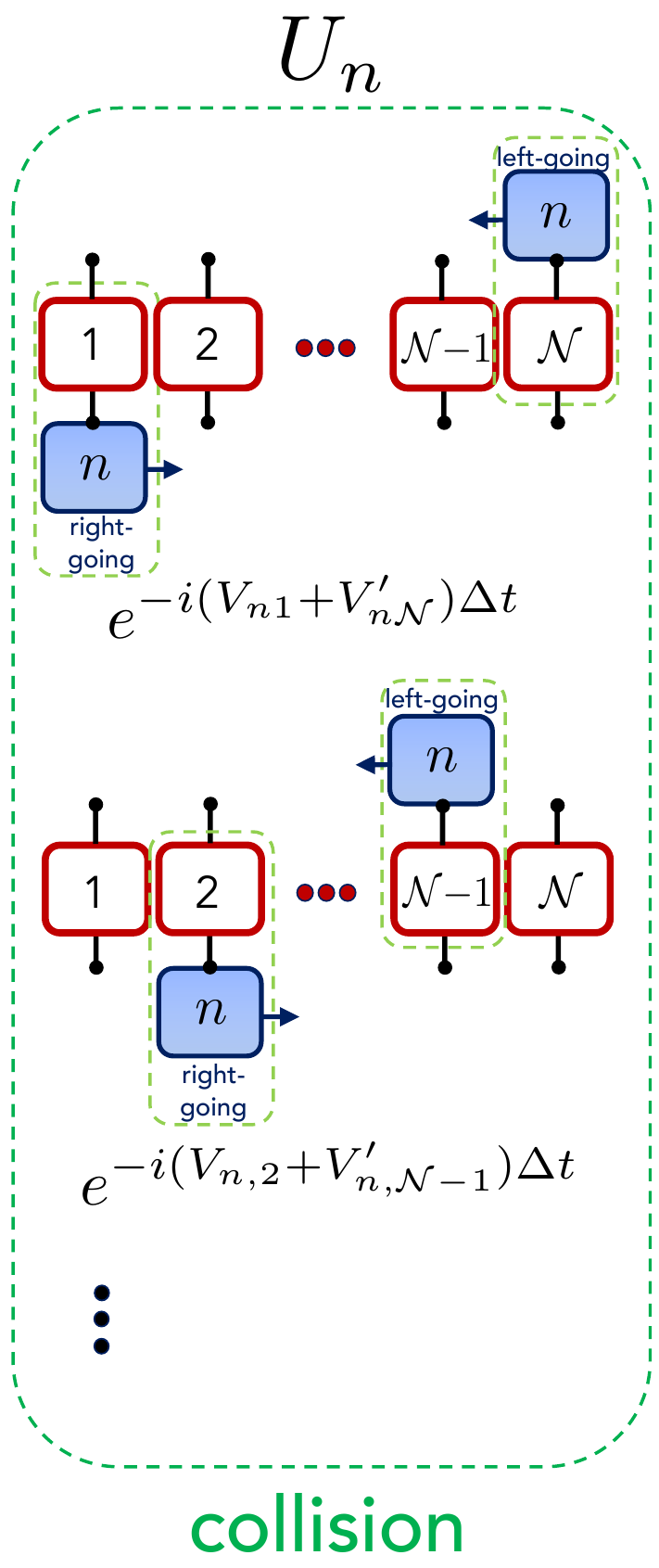}
	\caption{Collision unitary $U_n$'s decomposition into cascaded sub-collisions for a bidirectional waveguide [\cf\eq\eqref{deco3}]. In the $\nu$th sub-collision, the right- and left-going time bins respectively interact with coupling points $\nu$ and ${\cal N}-\nu$ (time grows from top to bottom).}\label{sub2}
\end{figure}

In Appendix \ref{appC}, we use \eqref{deco3} to show that a giant atom does not appear in $H_{\rm eff}$ whenever it is untangled from other atoms, i.e., when no coupling points of other atoms lie in between its two coupling points.

\section{More than two coupling points} \label{sec-multi}

The discussion in Sec.~\ref{sec-mec} in many respects relied on the property that a single giant atom with a $\pi$-phase shift between its two coupling points (i.e., the DF condition) fully decouples from the field, i.e., $U_n=\openone$ [\cf\eq\eqref{id}]. For more than two coupling points, the DF condition for a single giant atom does not necessarily entail $U_n= \openone$. The simplest example to see this is a single giant atom with three coupling points (${\cal N}\equiv {\cal N}_1=3$). The DF condition \eqref{cond-g2} occurs for (we drop subscript $j$ since there is only one atom; also we set $\varphi_1=0$) 
\begin{equation}
\varphi_2=\tfrac{2 \pi}{3} + 2 n \pi,~\varphi_3=\tfrac{4 \pi}{3} + 2 m \pi
\end{equation}
with $n,\,m$ integers.
Plugging these into the effective Hamiltonian \eqref{Heff2} for $\gamma'=0$ we get
\begin{align}
H_{\rm eff} =&\tfrac{\gamma}{2} \left(2\sin(\tfrac{2 \pi}{3}) + \sin(\tfrac{4 \pi}{3})\right)\sigma_{z}\neq 0\,
\end{align}
(the sum of the three sines is $\simeq 0.87$).

More generally, for an atom $j$ such that $x_{j'\neq j,\ell'}\not\in [x_{j,1},x_{j,{\cal N}_j}]$ and fulfilling the DF condition $\sum_{\ell} V_{n,\nu_{j,\ell}}=0$, in general
\begin{equation}
e^{-i V_{n,\nu_{j,{\cal N}_j}}\!\Delta t}\!\dots \,e^{-i V_{n,\nu_{j,1}}\Delta t}\neq \openone\,,\label{id2}
\end{equation}
where ${\cal N}_j>2$ (if ${\cal N}_j=2$, the identity holds). However, \eqref{id2} is anyway of the form $e^{-i \delta _j\sigma_z \Delta t}$ (with $\delta_j$ a frequency shift), hence all  terms of $H_{\rm eff}$ coupling $j$ to any other atom will vanish, i.e., in \eqref{Heff2} $J_{jj'}\neq 0$ only for $j'=j$. Thus, if the only focus is coupling the atoms, then the braided topology remains the only one yielding a non-trivial $H_{\rm eff}$. This remains true for a chiral waveguide ($\gamma'\neq 0$) since \eqref{jjj2} shows that if $J_{jj'}= 0$ for $\gamma'=0$ then it vanishes also for $\gamma'\neq 0$.

\section{Conclusions}\label{sec-conc}

In this work, we investigated the physical mechanism underpinning implementation of DF Hamiltonians with giant atoms. We first introduced a general framework for obtaining DF Hamiltonians through second-order interactions mediated by an environment. The key ``DF condition" is having an interaction Hamiltonian averaging to zero over a suitable coarse-grained time scale. The framework was first illustrated with standard dispersive Hamiltonians, in which case large detunings ensure a vanishing average interaction. We then considered giant atoms in a broadband waveguide and showed that, thanks to the non-local nature of the coupling, the DF condition can still be fulfilled, but in a qualitatively different way.

The above framework was then connected to a collisional picture of the joint dynamics of giant atoms and field in terms of elementary pairwise collisions between the atoms and field time bins. We showed that each collision can be decomposed as cascaded subcollisions, providing an intuitive understanding of the origin of the effective Hamiltonian. This was used to interpret the relationship between topology of the coupling points and occurrence of trivial/non-trivial DF Hamiltonians. In addition, we showed that the giant atoms dynamics can be mapped into a system shuttling between the atoms and subject to periodic phase kicks so as to effectively mediate a DF inter-atomic interaction, a mechanism in some respects reminiscent of dynamical decoupling schemes.

While here we did not consider lossy photonic environments \cite{MolmerPRA02,SorensenPRA11, Hammerer2019}, the considered framework could be naturally extended to accommodate these.
Likewise, a generalization to giant atoms in gapped structured reservoirs \cite{TudelaGiant19} appears viable.

On a methodological ground, we note that, although the collisional picture \cite{CP} and input-output/SLH formalism \cite{LalumierePRA13,CombesAdvPhyX17} are equivalent descriptions (since the underlying microscopic model and approximations are identical), the aspects of giant-atoms dynamics which this work focused on are best tackled through the former approach. This is essentially due to the decomposition into two-body unitaries, the hallmark of the collisional description. In this respect, the present work in particular showcases a type of problem where this approach is particularly advantageous.

From the specific viewpoint of collision models and their quantum information processing applications (see, e.g., \cite{suttonManipulating2015,laydenUniversal2016,CampbellPRA19}), Sections \ref{CM-def} and \ref{sec-qubit} in fact introduce a class of cascaded collision models implementing maximally-entangling multi-qubit gates free from decoherence. Notably, these correspond to {\it second-order} effective Hamiltonians, at variance with schemes in \rrefs \cite{suttonManipulating2015,laydenUniversal2016}  which are of first order (see also \rref\cite{altamiranoUnitarity2017}).

\section*{Acknowledgements}
We would like to thank G. M. Palma, S. Lorenzo, G. Falci and A. F. Kochum  for fruitful discussions. We acknowledge support from MIUR through project PRIN Project 2017SRN-BRK QUSHIP. A.C. acknowledges support from the Government of the Russian Federation through Agreement No. 074-02-2018-330 (2).

\bigskip
\appendix

\section{Proof of property}\label{app-proof}

The propagator in each time interval \eqref{Un-app} for $\overline V_n =0$ reduces to $ U_n=\openone -i  {\cal H}_{n}\Delta t$. Accordingly, 
\begin{equation}
\sigma_n=\sigma_{n-1}-i\,[ {\cal H}_{n}, \sigma_{n-1}]\Delta t\,.
\end{equation}
At the first step, $\sigma_{n-1}=\rho_0 \otimes \rho_E$, thus using condition \eqref{cond2} 
\begin{align}
	\sigma_{1}&=\rho_0 \otimes \rho_E-i\,[ {\cal H}_{n}, \rho_0 \otimes\rho_E]\,\Delta t\nonumber\\
	&=\rho_0 \otimes\rho_E-i\,[ {\cal H}_{n}, \rho_0\otimes\openone] \,(\openone\otimes\rho_E)\,\Delta t\nonumber\\
	&=\left(\rho_0 \,{-}i\,[ { H}_{\rm eff}, \rho_0]\right)\otimes \rho_E\Delta t\,
\end{align}
This implies $\sigma_{1}=\rho_1\otimes \rho_E$ with
\begin{equation}
\frac{\Delta  \rho_1}{\Delta t}=-i [H_{\rm eff},\rho_0]\,
\end{equation}
[recall \eq\eqref{Heff}]. By induction, we get that at each step $\sigma_{n}=\rho_n\otimes \rho_E$ with $\rho_n$ fulfilling 
\begin{equation}
\frac{\Delta  \rho_n}{\Delta t}=-i [H_{\rm eff},\rho_{n-1}]\,.\label{drho}
\end{equation}
Taking the continuous-time limit, $\rho_{n-1}\rightarrow \rho_t$, ${\Delta  \rho_n}/{\Delta t}\rightarrow \dot \rho$ so that \eqref{drho} reduces to \eq\eqref{evol}.

\section{$\,\,\tilde{{\cal H}}_n={\cal H}_n$}\label{AppA}

Let us begin with the unidirectional case. Each commutator in \eqref{H2bis} is explicitly worked out as
\begin{equation}\label{critical}
\left[V_{n\nu'},V_{n\nu}\right]
=\!\tfrac{\gamma}{\Delta t}\!\left(\left[ {S}_{\nu'}, {S}_{\nu}^\dag \right]\!{-}{\rm H.c.}\right) b_{n}^\dag b_{n}{+}\tfrac{\gamma}{\Delta t}\!\left({S}_{\nu'}^\dag{S}_{\nu}\! {-}{\rm H.c.}\right)\,.
\end{equation}
Upon comparison with \eqref{Hvac-rl} for $\gamma'=0$, the proof thus reduces to showing that the sum over $\nu{>}\nu'$ of terms $\propto b_n^\dag b_n$ vanish.

Each commutator $\left[ {S}_{\nu'}, {S}_{\nu}^\dag \right]$ is non-zero only when coupling points $\nu$ and $\nu'$ belong to the same atom. Thus, in light of \eqref{Snu},
\begin{align}
\sum_{\nu>\nu'}\left[S_{\nu},S_{\nu'}^\dag\right]-\rm{H.c.}&=\sum_j\sum_{\ell>\ell'}e^{i(\varphi_{j\ell}-\varphi_{j\ell'})}\left[\sigma_j,\sigma_j^\dag\right]-\rm{H.c.}\nonumber\\
&=\sum_j\left(\sum_{\ell>\ell'}e^{i(\varphi_{j\ell}-\varphi_{j\ell'})}{-}\rm{c.c.}\right)\!\sigma_{jz}\label{expr} 
\end{align}
(recall that $x_{j1}<x_{j2}<...$ ).
When \eqref{cond-g2} holds, the coefficient of $\sigma_{jz}$vanishes for each $j$ 
\begin{align}
\!\!\sum_{\ell>\ell'}e^{i(\varphi_{j\ell}-\varphi_{j\ell'})}-\rm{ c.c.}&=\sum_{\ell\ge\ell'}e^{i(\varphi_{j\ell}-\varphi_{j\ell'})}-\rm{ c.c.}\nonumber\\
&=\sum_{\ell=\ell'}^{\mathcal{N}_j}e^{i\varphi_{j\ell}}\sum_{\ell'=1}^{\mathcal{N}_j}e^{-i\varphi_{j\ell'}}-\rm{ c.c.}=0\,.\label{reas}
\end{align}
Thus 
\begin{equation}
\sum_{\nu>\nu'}\left[V_{n\nu'},V_{n\nu}\right]
=\tfrac{\gamma}{\Delta t}\sum_{\nu>\nu'}\left({S}_{\nu'}^\dag{S}_{\nu}\! -{\rm H.c.}\right)\,,
\end{equation}
completing the proof.

In the bidirectional case, each commutator in \eqref{comm-2way} reads
\begin{align}
&\left[V_{n,\nu'}+V'_{n,\,{\cal N}+1-\nu'},V_{n,\nu}+V'_{n,\,{\cal N}+1-\nu}\right]=
\nonumber\\&\,\,\,\,\,\,\,\,\,\,\,\,\,\,\,\,\,\,\,\left[V_{n,\nu'},V_{n,\nu}\right]+
\left[V'_{n,\,{\cal N}+1-\nu'},V'_{n,\,{\cal N}+1-\nu}\right] \nonumber
\nonumber\\&
\,\,\,\,\,\,\,\,\,\,\,\,\,\,\,\,\,\,\,+\left[V_{n,\nu'},V'_{n,\,{\cal N}+1-\nu}\right]+
\left[V'_{n,\,{\cal N}+1-\nu'},V_{n,\nu}\right].\label{VV2}
\end{align}
The last line features terms $\propto\! [b_n^\dag, b'_n]$ and $\propto \![{b'}_n^\dag, b_n]$, which vanish because left- and right-going time-bin operators commute. Additionally, 
there are terms $\propto b_n^\dag b'_n$ (or $\propto b_n^{'\dag} b_n$) featuring quantities like \eqref{expr} where however one of the two phases is primed: these vanish as well since \eqref{reas} holds even if $\varphi_{j\ell'}\rightarrow \varphi'_{j\ell'}$.
We are thus only left with terms analogous to \eqref{critical} given by
\begin{align}
& \left[V_{n,\nu'},V_{n,\nu}\right]+
\left[V'_{n,\,{\cal N}+1-\nu'},V'_{n,\,{\cal N}+1-\nu}\right] =\nonumber\\&
\,\,\,\,\,\,\,\,\,\,\,\,\tfrac{\gamma}{\Delta t}\!\left({S}_{\nu'}^\dag{S}_{\nu}\! + \! {S'}_{{\cal N}+1-\nu'}^\dag{S'}_{{\cal N}+1-\nu}\! -{\rm H.c.}\right)\,.
\end{align}
Summing this over $\nu>\nu'$ yields $\tfrac{\gamma}{\Delta t}\!\left({S}_{\nu'}^\dag{S}_{\nu}\! + \! {S'}_{\nu}^\dag{S'}_{\nu'}\! -{\rm H.c.}\right)$ (where we used that ${\cal N}{+}1{-}\nu'>{\cal N}{+}1{-}\nu$ for $\nu>\nu'$), completing the proof.

\section{Extension of Section\ref{many} to a birectional waveguide}\label{appC}

In order to extend the considerations in Sec.~\ref{many}  to a bidirectional waveguide, we essentially need to show that, for an atom $j$ such that $x_{j',\ell'}\not\in [x_{j,1},x_{j,2}]$ for any $j'\neq j$ and $\ell'=1,2$ (recall Sec.~\ref{many}), $\sigma_{j}$ and $\sigma_{j}^\dag$ do not appear in $H_{\rm eff}$ under the DF condition. We first recall that $\overline{V}_n=0\Leftrightarrow \overline{V'}_n=0$ (since $S=0\Leftrightarrow S'=0$). In this configuration, the overall coupling point index runs over $\nu=1,2,...,k_j,k_j+1,...,{\cal N}$ with $k_j\leftrightarrow (j,1)$ and $k_j+1\leftrightarrow (j,2)$ labeling the left and right coupling points of atom $j$. Accordingly, in unitary \eqref{deco3}, the sub-collision unitaries involving the $j$th atom are (to make notation lighter we drop subscript $n$)
\begin{eqnarray}
\tilde U_j=e^{-i  (V_{k_j+1}+V'_{{\cal N}-k_j}) \Delta t} e^{-i  (V_{k_j}+V'_{{\cal N}+1-k_j}) \Delta t}\label{exp1}
\end{eqnarray}
and 
\begin{eqnarray}
\tilde W_j=e^{-i  (V_{{\cal N}+1-k_j}+V'_{k_j}) \Delta t} e^{-i  (V_{{\cal N}-k_j}+V'_{k_j+1}) \Delta t}\,.\label{exp2}
\end{eqnarray}
Upon inspection, the pair of $V$'s in each exponent commute because one involves a coupling point of atom $j$ and a right-going time bin, while the other one features a coupling point of an atom $j'\neq j$ and a left-going time bin. Thereby, \eqref{exp1} can be decomposed as
\begin{eqnarray}
\tilde U_j&=&e^{-i  V_{k_j+1} \Delta t} e^{-i  V'_{{\cal N}-k_j} \Delta t} e^{-i  V_{k_j} \Delta t}e^{-i  V'_{{\cal N}+1-k_j} \Delta t}\nonumber\\
&=&e^{-i  V_{k_j+1} \Delta t} e^{-i  V_{k_j} \Delta t} e^{-i  V'_{{\cal N}-k_j} \Delta t} e^{-i  V'_{{\cal N}+1-k_j} \Delta t}\,.\label{exp3}
\end{eqnarray}
Under the DF condition, $V_{k_j+1}=-V_{k_j}$ so that the first two exponentials in the last line reduce to the identity. An analogous conclusion holds for \eqref{exp2}. We thus conclude that $U_n$, hence $H_{\rm eff}$, does not contain atom $j$, completing the proof.

\bibliography{WQEDCCC}

\begin{thebibliography}{59}%
\makeatletter
\providecommand \@ifxundefined [1]{%
 \@ifx{#1\undefined}
}%
\providecommand \@ifnum [1]{%
 \ifnum #1\expandafter \@firstoftwo
 \else \expandafter \@secondoftwo
 \fi
}%
\providecommand \@ifx [1]{%
 \ifx #1\expandafter \@firstoftwo
 \else \expandafter \@secondoftwo
 \fi
}%
\providecommand \natexlab [1]{#1}%
\providecommand \enquote  [1]{``#1''}%
\providecommand \bibnamefont  [1]{#1}%
\providecommand \bibfnamefont [1]{#1}%
\providecommand \citenamefont [1]{#1}%
\providecommand \href@noop [0]{\@secondoftwo}%
\providecommand \href [0]{\begingroup \@sanitize@url \@href}%
\providecommand \@href[1]{\@@startlink{#1}\@@href}%
\providecommand \@@href[1]{\endgroup#1\@@endlink}%
\providecommand \@sanitize@url [0]{\catcode `\\12\catcode `\$12\catcode
  `\&12\catcode `\#12\catcode `\^12\catcode `\_12\catcode `\%12\relax}%
\providecommand \@@startlink[1]{}%
\providecommand \@@endlink[0]{}%
\providecommand \url  [0]{\begingroup\@sanitize@url \@url }%
\providecommand \@url [1]{\endgroup\@href {#1}{\urlprefix }}%
\providecommand \urlprefix  [0]{URL }%
\providecommand \Eprint [0]{\href }%
\providecommand \doibase [0]{http://dx.doi.org/}%
\providecommand \selectlanguage [0]{\@gobble}%
\providecommand \bibinfo  [0]{\@secondoftwo}%
\providecommand \bibfield  [0]{\@secondoftwo}%
\providecommand \translation [1]{[#1]}%
\providecommand \BibitemOpen [0]{}%
\providecommand \bibitemStop [0]{}%
\providecommand \bibitemNoStop [0]{.\EOS\space}%
\providecommand \EOS [0]{\spacefactor3000\relax}%
\providecommand \BibitemShut  [1]{\csname bibitem#1\endcsname}%
\let\auto@bib@innerbib\@empty
\bibitem [{\citenamefont {S\o{}rensen}\ and\ \citenamefont
  {M\o{}lmer}(1999)}]{MolmerPRL99}%
  \BibitemOpen
  \bibfield  {author} {\bibinfo {author} {\bibfnamefont {Anders}\ \bibnamefont
  {S\o{}rensen}}\ and\ \bibinfo {author} {\bibfnamefont {Klaus}\ \bibnamefont
  {M\o{}lmer}},\ }\bibfield  {title} {\enquote {\bibinfo {title} {Quantum
  computation with ions in thermal motion},}\ }\href {\doibase
  10.1103/PhysRevLett.82.1971} {\bibfield  {journal} {\bibinfo  {journal}
  {Phys. Rev. Lett.}\ }\textbf {\bibinfo {volume} {82}},\ \bibinfo {pages}
  {1971--1974} (\bibinfo {year} {1999})}\BibitemShut {NoStop}%
\bibitem [{\citenamefont {Beige}\ \emph {et~al.}(2000)\citenamefont {Beige},
  \citenamefont {Braun}, \citenamefont {Tregenna},\ and\ \citenamefont
  {Knight}}]{BeigePRL00}%
  \BibitemOpen
  \bibfield  {author} {\bibinfo {author} {\bibfnamefont {Almut}\ \bibnamefont
  {Beige}}, \bibinfo {author} {\bibfnamefont {Daniel}\ \bibnamefont {Braun}},
  \bibinfo {author} {\bibfnamefont {Ben}\ \bibnamefont {Tregenna}}, \ and\
  \bibinfo {author} {\bibfnamefont {Peter~L.}\ \bibnamefont {Knight}},\
  }\bibfield  {title} {\enquote {\bibinfo {title} {Quantum computing using
  dissipation to remain in a decoherence-free subspace},}\ }\href {\doibase
  10.1103/PhysRevLett.85.1762} {\bibfield  {journal} {\bibinfo  {journal}
  {Phys. Rev. Lett.}\ }\textbf {\bibinfo {volume} {85}},\ \bibinfo {pages}
  {1762--1765} (\bibinfo {year} {2000})}\BibitemShut {NoStop}%
\bibitem [{\citenamefont {Kempe}\ \emph {et~al.}(2001)\citenamefont {Kempe},
  \citenamefont {Bacon}, \citenamefont {Lidar},\ and\ \citenamefont
  {Whaley}}]{KempePRA01}%
  \BibitemOpen
  \bibfield  {author} {\bibinfo {author} {\bibfnamefont {J.}~\bibnamefont
  {Kempe}}, \bibinfo {author} {\bibfnamefont {D.}~\bibnamefont {Bacon}},
  \bibinfo {author} {\bibfnamefont {D.~A.}\ \bibnamefont {Lidar}}, \ and\
  \bibinfo {author} {\bibfnamefont {K.~B.}\ \bibnamefont {Whaley}},\ }\bibfield
   {title} {\enquote {\bibinfo {title} {Theory of decoherence-free
  fault-tolerant universal quantum computation},}\ }\href {\doibase
  10.1103/PhysRevA.63.042307} {\bibfield  {journal} {\bibinfo  {journal} {Phys.
  Rev. A}\ }\textbf {\bibinfo {volume} {63}},\ \bibinfo {pages} {042307}
  (\bibinfo {year} {2001})}\BibitemShut {NoStop}%
\bibitem [{\citenamefont {Facchi}\ and\ \citenamefont
  {Pascazio}(2008)}]{FacchiJPA2008}%
  \BibitemOpen
  \bibfield  {author} {\bibinfo {author} {\bibfnamefont {P}~\bibnamefont
  {Facchi}}\ and\ \bibinfo {author} {\bibfnamefont {S}~\bibnamefont
  {Pascazio}},\ }\bibfield  {title} {\enquote {\bibinfo {title} {Quantum zeno
  dynamics: mathematical and physical aspects},}\ }\href {\doibase
  10.1088/1751-8113/41/49/493001} {\bibfield  {journal} {\bibinfo  {journal}
  {Journal of Physics A: Mathematical and Theoretical}\ }\textbf {\bibinfo
  {volume} {41}},\ \bibinfo {pages} {493001} (\bibinfo {year}
  {2008})}\BibitemShut {NoStop}%
\bibitem [{\citenamefont {Shahmoon}\ and\ \citenamefont
  {Kurizki}(2013)}]{Shahmoon2013}%
  \BibitemOpen
  \bibfield  {author} {\bibinfo {author} {\bibfnamefont {Ephraim}\ \bibnamefont
  {Shahmoon}}\ and\ \bibinfo {author} {\bibfnamefont {Gershon}\ \bibnamefont
  {Kurizki}},\ }\bibfield  {title} {\enquote {\bibinfo {title} {{Nonradiative
  interaction and entanglement between distant atoms}},}\ }\href {\doibase
  10.1103/PhysRevA.87.033831} {\bibfield  {journal} {\bibinfo  {journal}
  {Physical Review A - Atomic, Molecular, and Optical Physics}\ }\textbf
  {\bibinfo {volume} {87}},\ \bibinfo {pages} {033831} (\bibinfo {year}
  {2013})}\BibitemShut {NoStop}%
\bibitem [{\citenamefont {Douglas}\ \emph {et~al.}(2015)\citenamefont
  {Douglas}, \citenamefont {Habibian}, \citenamefont {Hung}, \citenamefont
  {Gorshkov}, \citenamefont {Kimble},\ and\ \citenamefont
  {Chang}}]{Douglas2015b}%
  \BibitemOpen
  \bibfield  {author} {\bibinfo {author} {\bibfnamefont {J.~S.}\ \bibnamefont
  {Douglas}}, \bibinfo {author} {\bibfnamefont {H.}~\bibnamefont {Habibian}},
  \bibinfo {author} {\bibfnamefont {C.~L.}\ \bibnamefont {Hung}}, \bibinfo
  {author} {\bibfnamefont {A.~V.}\ \bibnamefont {Gorshkov}}, \bibinfo {author}
  {\bibfnamefont {H.~J.}\ \bibnamefont {Kimble}}, \ and\ \bibinfo {author}
  {\bibfnamefont {D.~E.}\ \bibnamefont {Chang}},\ }\bibfield  {title} {\enquote
  {\bibinfo {title} {{Quantum many-body models with cold atoms coupled to
  photonic crystals}},}\ }\href {\doibase 10.1038/nphoton.2015.57} {\bibfield
  {journal} {\bibinfo  {journal} {Nature Photonics}\ }\textbf {\bibinfo
  {volume} {9}},\ \bibinfo {pages} {326--331} (\bibinfo {year} {2015})},\
  \Eprint {http://arxiv.org/abs/1312.2435} {1312.2435} \BibitemShut {NoStop}%
\bibitem [{\citenamefont {Paulisch}\ \emph {et~al.}(2016)\citenamefont
  {Paulisch}, \citenamefont {Kimble},\ and\ \citenamefont
  {Gonz{\'{a}}lez-Tudela}}]{PaulischNJP2016}%
  \BibitemOpen
  \bibfield  {author} {\bibinfo {author} {\bibfnamefont {V}~\bibnamefont
  {Paulisch}}, \bibinfo {author} {\bibfnamefont {H~J}\ \bibnamefont {Kimble}},
  \ and\ \bibinfo {author} {\bibfnamefont {A}~\bibnamefont
  {Gonz{\'{a}}lez-Tudela}},\ }\bibfield  {title} {\enquote {\bibinfo {title}
  {Universal quantum computation in waveguide {QED} using decoherence free
  subspaces},}\ }\href {\doibase 10.1088/1367-2630/18/4/043041} {\bibfield
  {journal} {\bibinfo  {journal} {New Journal of Physics}\ }\textbf {\bibinfo
  {volume} {18}},\ \bibinfo {pages} {043041} (\bibinfo {year}
  {2016})}\BibitemShut {NoStop}%
\bibitem [{\citenamefont {Cohen-Tannoudji}\ \emph {et~al.}(2004)\citenamefont
  {Cohen-Tannoudji}, \citenamefont {Dupont-Roc}, \citenamefont {Grynberg},\
  and\ \citenamefont {Thickstun}}]{CohenAP}%
  \BibitemOpen
  \bibfield  {author} {\bibinfo {author} {\bibfnamefont {C.}~\bibnamefont
  {Cohen-Tannoudji}}, \bibinfo {author} {\bibfnamefont {J.}~\bibnamefont
  {Dupont-Roc}}, \bibinfo {author} {\bibfnamefont {G.}~\bibnamefont
  {Grynberg}}, \ and\ \bibinfo {author} {\bibfnamefont {P.}~\bibnamefont
  {Thickstun}},\ }\href@noop {} {\emph {\bibinfo {title} {Atom-photon
  interactions: basic processes and applications}}}\ (\bibinfo  {publisher}
  {Wiley Online Library, 1992},\ \bibinfo {year} {2004})\BibitemShut {NoStop}%
\bibitem [{\citenamefont {Breuer}(2012)}]{breuerFoundations2012}%
  \BibitemOpen
  \bibfield  {author} {\bibinfo {author} {\bibfnamefont {Heinz~Peter}\
  \bibnamefont {Breuer}},\ }\bibfield  {title} {\enquote {\bibinfo {title}
  {{Foundations and measures of quantum non-Markovianity}},}\ }\href {\doibase
  10.1088/0953-4075/45/15/154001} {\bibfield  {journal} {\bibinfo  {journal}
  {Journal of Physics B: Atomic, Molecular and Optical Physics}\ }\textbf
  {\bibinfo {volume} {45}},\ \bibinfo {pages} {154001} (\bibinfo {year}
  {2012})},\ \Eprint {http://arxiv.org/abs/1206.5346} {1206.5346} \BibitemShut
  {NoStop}%
\bibitem [{\citenamefont {Zheng}\ and\ \citenamefont
  {Guo}(2000)}]{ZhengPRL2000}%
  \BibitemOpen
  \bibfield  {author} {\bibinfo {author} {\bibfnamefont {Shi-Biao}\
  \bibnamefont {Zheng}}\ and\ \bibinfo {author} {\bibfnamefont {Guang-Can}\
  \bibnamefont {Guo}},\ }\bibfield  {title} {\enquote {\bibinfo {title}
  {Efficient scheme for two-atom entanglement and quantum information
  processing in cavity qed},}\ }\href {\doibase 10.1103/PhysRevLett.85.2392}
  {\bibfield  {journal} {\bibinfo  {journal} {Phys. Rev. Lett.}\ }\textbf
  {\bibinfo {volume} {85}},\ \bibinfo {pages} {2392--2395} (\bibinfo {year}
  {2000})}\BibitemShut {NoStop}%
\bibitem [{\citenamefont {Majer}\ \emph {et~al.}(2007)\citenamefont {Majer},
  \citenamefont {Chow}, \citenamefont {Gambetta}, \citenamefont {Koch},
  \citenamefont {Johnson}, \citenamefont {Schreier}, \citenamefont {Frunzio},
  \citenamefont {Schuster}, \citenamefont {Houck}, \citenamefont {Wallraff},
  \citenamefont {Blais}, \citenamefont {Devoret}, \citenamefont {Girvin},\ and\
  \citenamefont {Schoelkopf}}]{MajerNat07}%
  \BibitemOpen
  \bibfield  {author} {\bibinfo {author} {\bibfnamefont {J.}~\bibnamefont
  {Majer}}, \bibinfo {author} {\bibfnamefont {J.~M.}\ \bibnamefont {Chow}},
  \bibinfo {author} {\bibfnamefont {J.~M.}\ \bibnamefont {Gambetta}}, \bibinfo
  {author} {\bibfnamefont {Jens}\ \bibnamefont {Koch}}, \bibinfo {author}
  {\bibfnamefont {B.~R.}\ \bibnamefont {Johnson}}, \bibinfo {author}
  {\bibfnamefont {J.~A.}\ \bibnamefont {Schreier}}, \bibinfo {author}
  {\bibfnamefont {L.}~\bibnamefont {Frunzio}}, \bibinfo {author} {\bibfnamefont
  {D.~I.}\ \bibnamefont {Schuster}}, \bibinfo {author} {\bibfnamefont {A.~A.}\
  \bibnamefont {Houck}}, \bibinfo {author} {\bibfnamefont {A.}~\bibnamefont
  {Wallraff}}, \bibinfo {author} {\bibfnamefont {A.}~\bibnamefont {Blais}},
  \bibinfo {author} {\bibfnamefont {M.~H.}\ \bibnamefont {Devoret}}, \bibinfo
  {author} {\bibfnamefont {S.~M.}\ \bibnamefont {Girvin}}, \ and\ \bibinfo
  {author} {\bibfnamefont {R.~J.}\ \bibnamefont {Schoelkopf}},\ }\bibfield
  {title} {\enquote {\bibinfo {title} {{Coupling superconducting qubits via a
  cavity bus}},}\ }\href {\doibase 10.1038/nature06184} {\bibfield  {journal}
  {\bibinfo  {journal} {Nature}\ }\textbf {\bibinfo {volume} {449}},\ \bibinfo
  {pages} {443--447} (\bibinfo {year} {2007})},\ \Eprint
  {http://arxiv.org/abs/0709.2135} {0709.2135} \BibitemShut {NoStop}%
\bibitem [{\citenamefont {Ogden}\ \emph {et~al.}(2008)\citenamefont {Ogden},
  \citenamefont {Irish},\ and\ \citenamefont {Kim}}]{IrishPRA08}%
  \BibitemOpen
  \bibfield  {author} {\bibinfo {author} {\bibfnamefont {C.~D.}\ \bibnamefont
  {Ogden}}, \bibinfo {author} {\bibfnamefont {E.~K.}\ \bibnamefont {Irish}}, \
  and\ \bibinfo {author} {\bibfnamefont {M.~S.}\ \bibnamefont {Kim}},\
  }\bibfield  {title} {\enquote {\bibinfo {title} {Dynamics in a coupled-cavity
  array},}\ }\href {\doibase 10.1103/PhysRevA.78.063805} {\bibfield  {journal}
  {\bibinfo  {journal} {Phys. Rev. A}\ }\textbf {\bibinfo {volume} {78}},\
  \bibinfo {pages} {063805} (\bibinfo {year} {2008})}\BibitemShut {NoStop}%
\bibitem [{\citenamefont {Chang}\ \emph {et~al.}(2018)\citenamefont {Chang},
  \citenamefont {Douglas}, \citenamefont {Gonz\'alez-Tudela}, \citenamefont
  {Hung},\ and\ \citenamefont {Kimble}}]{TudelaRMP18}%
  \BibitemOpen
  \bibfield  {author} {\bibinfo {author} {\bibfnamefont {D.~E.}\ \bibnamefont
  {Chang}}, \bibinfo {author} {\bibfnamefont {J.~S.}\ \bibnamefont {Douglas}},
  \bibinfo {author} {\bibfnamefont {A.}~\bibnamefont {Gonz\'alez-Tudela}},
  \bibinfo {author} {\bibfnamefont {C.-L.}\ \bibnamefont {Hung}}, \ and\
  \bibinfo {author} {\bibfnamefont {H.~J.}\ \bibnamefont {Kimble}},\ }\bibfield
   {title} {\enquote {\bibinfo {title} {Colloquium: Quantum matter built from
  nanoscopic lattices of atoms and photons},}\ }\href {\doibase
  10.1103/RevModPhys.90.031002} {\bibfield  {journal} {\bibinfo  {journal}
  {Rev. Mod. Phys.}\ }\textbf {\bibinfo {volume} {90}},\ \bibinfo {pages}
  {031002} (\bibinfo {year} {2018})}\BibitemShut {NoStop}%
\bibitem [{\citenamefont {Kockum}\ \emph {et~al.}(2018)\citenamefont {Kockum},
  \citenamefont {Johansson},\ and\ \citenamefont {Nori}}]{KockumPRL2018}%
  \BibitemOpen
  \bibfield  {author} {\bibinfo {author} {\bibfnamefont {Anton~Frisk}\
  \bibnamefont {Kockum}}, \bibinfo {author} {\bibfnamefont {G\"oran}\
  \bibnamefont {Johansson}}, \ and\ \bibinfo {author} {\bibfnamefont {Franco}\
  \bibnamefont {Nori}},\ }\bibfield  {title} {\enquote {\bibinfo {title}
  {Decoherence-free interaction between giant atoms in waveguide quantum
  electrodynamics},}\ }\href {\doibase 10.1103/PhysRevLett.120.140404}
  {\bibfield  {journal} {\bibinfo  {journal} {Phys. Rev. Lett.}\ }\textbf
  {\bibinfo {volume} {120}},\ \bibinfo {pages} {140404} (\bibinfo {year}
  {2018})}\BibitemShut {NoStop}%
\bibitem [{\citenamefont {Kannan}\ \emph {et~al.}(2019)\citenamefont {Kannan},
  \citenamefont {Ruckriegel}, \citenamefont {Campbell}, \citenamefont {Kockum},
  \citenamefont {Braum{\"u}ller}, \citenamefont {Kim}, \citenamefont
  {Kjaergaard}, \citenamefont {Krantz}, \citenamefont {Melville}, \citenamefont
  {Niedzielski} \emph {et~al.}}]{OliverGiant2019}%
  \BibitemOpen
  \bibfield  {author} {\bibinfo {author} {\bibfnamefont {Bharath}\ \bibnamefont
  {Kannan}}, \bibinfo {author} {\bibfnamefont {Max}\ \bibnamefont
  {Ruckriegel}}, \bibinfo {author} {\bibfnamefont {Daniel}\ \bibnamefont
  {Campbell}}, \bibinfo {author} {\bibfnamefont {Anton~Frisk}\ \bibnamefont
  {Kockum}}, \bibinfo {author} {\bibfnamefont {Jochen}\ \bibnamefont
  {Braum{\"u}ller}}, \bibinfo {author} {\bibfnamefont {David}\ \bibnamefont
  {Kim}}, \bibinfo {author} {\bibfnamefont {Morten}\ \bibnamefont
  {Kjaergaard}}, \bibinfo {author} {\bibfnamefont {Philip}\ \bibnamefont
  {Krantz}}, \bibinfo {author} {\bibfnamefont {Alexander}\ \bibnamefont
  {Melville}}, \bibinfo {author} {\bibfnamefont {Bethany~M}\ \bibnamefont
  {Niedzielski}},  \emph {et~al.},\ }\bibfield  {title} {\enquote {\bibinfo
  {title} {Waveguide quantum electrodynamics with giant superconducting
  artificial atoms},}\ }\href {https://arxiv.org/abs/1912.12233} {\bibfield
  {journal} {\bibinfo  {journal} {arXiv preprint arXiv:1912.12233}\ } (\bibinfo
  {year} {2019})}\BibitemShut {NoStop}%
\bibitem [{\citenamefont {Kockum}(2019)}]{Kockum5years}%
  \BibitemOpen
  \bibfield  {author} {\bibinfo {author} {\bibfnamefont {Anton~Frisk}\
  \bibnamefont {Kockum}},\ }\bibfield  {title} {\enquote {\bibinfo {title}
  {Quantum optics with giant atoms--the first five years},}\ }\href
  {https://arxiv.org/abs/1912.13012} {\bibfield  {journal} {\bibinfo  {journal}
  {arXiv preprint arXiv:1912.13012}\ } (\bibinfo {year} {2019})}\BibitemShut
  {NoStop}%
\bibitem [{\citenamefont {{Frisk Kockum}}\ \emph {et~al.}(2014)\citenamefont
  {{Frisk Kockum}}, \citenamefont {Delsing},\ and\ \citenamefont
  {Johansson}}]{FriskKockumPRA14}%
  \BibitemOpen
  \bibfield  {author} {\bibinfo {author} {\bibfnamefont {Anton}\ \bibnamefont
  {{Frisk Kockum}}}, \bibinfo {author} {\bibfnamefont {Per}\ \bibnamefont
  {Delsing}}, \ and\ \bibinfo {author} {\bibfnamefont {G{\"{o}}ran}\
  \bibnamefont {Johansson}},\ }\bibfield  {title} {\enquote {\bibinfo {title}
  {{Designing frequency-dependent relaxation rates and Lamb shifts for a giant
  artificial atom}},}\ }\href {\doibase 10.1103/PhysRevA.90.013837} {\bibfield
  {journal} {\bibinfo  {journal} {Physical Review A - Atomic, Molecular, and
  Optical Physics}\ }\textbf {\bibinfo {volume} {90}},\ \bibinfo {pages}
  {13837} (\bibinfo {year} {2014})},\ \Eprint {http://arxiv.org/abs/1406.0350}
  {1406.0350} \BibitemShut {NoStop}%
\bibitem [{\citenamefont {Guo}\ \emph {et~al.}(2017)\citenamefont {Guo},
  \citenamefont {Grimsmo}, \citenamefont {Kockum}, \citenamefont {Pletyukhov},\
  and\ \citenamefont {Johansson}}]{KockumPRA2017}%
  \BibitemOpen
  \bibfield  {author} {\bibinfo {author} {\bibfnamefont {Lingzhen}\
  \bibnamefont {Guo}}, \bibinfo {author} {\bibfnamefont {Arne}\ \bibnamefont
  {Grimsmo}}, \bibinfo {author} {\bibfnamefont {Anton~Frisk}\ \bibnamefont
  {Kockum}}, \bibinfo {author} {\bibfnamefont {Mikhail}\ \bibnamefont
  {Pletyukhov}}, \ and\ \bibinfo {author} {\bibfnamefont {G\"oran}\
  \bibnamefont {Johansson}},\ }\bibfield  {title} {\enquote {\bibinfo {title}
  {Giant acoustic atom: A single quantum system with a deterministic time
  delay},}\ }\href@noop {} {\bibfield  {journal} {\bibinfo  {journal} {Phys.
  Rev. A}\ }\textbf {\bibinfo {volume} {95}},\ \bibinfo {pages} {053821}
  (\bibinfo {year} {2017})}\BibitemShut {NoStop}%
\bibitem [{\citenamefont {Karg}\ \emph {et~al.}(2019)\citenamefont {Karg},
  \citenamefont {Gouraud}, \citenamefont {Treutlein},\ and\ \citenamefont
  {Hammerer}}]{Hammerer2019}%
  \BibitemOpen
  \bibfield  {author} {\bibinfo {author} {\bibfnamefont {Thomas~M.}\
  \bibnamefont {Karg}}, \bibinfo {author} {\bibfnamefont {Baptiste}\
  \bibnamefont {Gouraud}}, \bibinfo {author} {\bibfnamefont {Philipp}\
  \bibnamefont {Treutlein}}, \ and\ \bibinfo {author} {\bibfnamefont {Klemens}\
  \bibnamefont {Hammerer}},\ }\bibfield  {title} {\enquote {\bibinfo {title}
  {Remote hamiltonian interactions mediated by light},}\ }\href {\doibase
  10.1103/PhysRevA.99.063829} {\bibfield  {journal} {\bibinfo  {journal} {Phys.
  Rev. A}\ }\textbf {\bibinfo {volume} {99}},\ \bibinfo {pages} {063829}
  (\bibinfo {year} {2019})}\BibitemShut {NoStop}%
\bibitem [{\citenamefont {Andersson}\ \emph {et~al.}(2019)\citenamefont
  {Andersson}, \citenamefont {Suri}, \citenamefont {Guo}, \citenamefont
  {Aref},\ and\ \citenamefont {Delsing}}]{AnderssonarXiv18}%
  \BibitemOpen
  \bibfield  {author} {\bibinfo {author} {\bibfnamefont {Gustav}\ \bibnamefont
  {Andersson}}, \bibinfo {author} {\bibfnamefont {Baladitya}\ \bibnamefont
  {Suri}}, \bibinfo {author} {\bibfnamefont {Lingzhen}\ \bibnamefont {Guo}},
  \bibinfo {author} {\bibfnamefont {Thomas}\ \bibnamefont {Aref}}, \ and\
  \bibinfo {author} {\bibfnamefont {Per}\ \bibnamefont {Delsing}},\ }\href
  {\doibase 10.1038/s41567-019-0605-6} {\enquote {\bibinfo {title}
  {{Non-exponential decay of a giant artificial atom}},}\ } (\bibinfo {year}
  {2019}),\ \Eprint {http://arxiv.org/abs/1812.01302} {1812.01302} \BibitemShut
  {NoStop}%
\bibitem [{\citenamefont {Guimond}\ \emph {et~al.}(2020)\citenamefont
  {Guimond}, \citenamefont {Vermersch}, \citenamefont {Juan}, \citenamefont
  {Sharafiev}, \citenamefont {Kirchmair},\ and\ \citenamefont
  {Zoller}}]{guimond2020unidirectional}%
  \BibitemOpen
  \bibfield  {author} {\bibinfo {author} {\bibfnamefont {P-O}\ \bibnamefont
  {Guimond}}, \bibinfo {author} {\bibfnamefont {B}~\bibnamefont {Vermersch}},
  \bibinfo {author} {\bibfnamefont {ML}~\bibnamefont {Juan}}, \bibinfo {author}
  {\bibfnamefont {A}~\bibnamefont {Sharafiev}}, \bibinfo {author}
  {\bibfnamefont {G}~\bibnamefont {Kirchmair}}, \ and\ \bibinfo {author}
  {\bibfnamefont {P}~\bibnamefont {Zoller}},\ }\bibfield  {title} {\enquote
  {\bibinfo {title} {A unidirectional on-chip photonic interface for
  superconducting circuits},}\ }\href
  {https://doi.org/10.1038/s41534-020-0261-9} {\bibfield  {journal} {\bibinfo
  {journal} {npj Quantum Information}\ }\textbf {\bibinfo {volume} {6}},\
  \bibinfo {pages} {1--12} (\bibinfo {year} {2020})}\BibitemShut {NoStop}%
\bibitem [{\citenamefont {Guo}\ \emph {et~al.}(2020)\citenamefont {Guo},
  \citenamefont {Kockum}, \citenamefont {Marquardt},\ and\ \citenamefont
  {Johansson}}]{guo2019oscillating}%
  \BibitemOpen
  \bibfield  {author} {\bibinfo {author} {\bibfnamefont {Lingzhen}\
  \bibnamefont {Guo}}, \bibinfo {author} {\bibfnamefont {Anton~Frisk}\
  \bibnamefont {Kockum}}, \bibinfo {author} {\bibfnamefont {Florian}\
  \bibnamefont {Marquardt}}, \ and\ \bibinfo {author} {\bibfnamefont
  {G{\"o}ran}\ \bibnamefont {Johansson}},\ }\bibfield  {title} {\enquote
  {\bibinfo {title} {Oscillating bound states for a giant atom},}\ }\href
  {https://arxiv.org/abs/1911.13028} {\bibfield  {journal} {\bibinfo  {journal}
  {Phys. Rev. Research}\ }\textbf {\bibinfo {volume} {2}},\ \bibinfo {pages}
  {043014} (\bibinfo {year} {2020})}\BibitemShut {NoStop}%
\bibitem [{\citenamefont {Vadiraj}\ \emph {et~al.}(2020)\citenamefont
  {Vadiraj}, \citenamefont {Ask}, \citenamefont {McConkey}, \citenamefont
  {Nsanzineza}, \citenamefont {Sandbo~Chang}, \citenamefont {Kockum},\ and\
  \citenamefont {Wilson}}]{Wilson2020}%
  \BibitemOpen
  \bibfield  {author} {\bibinfo {author} {\bibfnamefont {A.~M.}\ \bibnamefont
  {Vadiraj}}, \bibinfo {author} {\bibfnamefont {Andreas}\ \bibnamefont {Ask}},
  \bibinfo {author} {\bibfnamefont {T.~G.}\ \bibnamefont {McConkey}}, \bibinfo
  {author} {\bibfnamefont {I.}~\bibnamefont {Nsanzineza}}, \bibinfo {author}
  {\bibfnamefont {C.~W.}\ \bibnamefont {Sandbo~Chang}}, \bibinfo {author}
  {\bibfnamefont {Anton~Frisk}\ \bibnamefont {Kockum}}, \ and\ \bibinfo
  {author} {\bibfnamefont {C.~M.}\ \bibnamefont {Wilson}},\ }\bibfield  {title}
  {\enquote {\bibinfo {title} {Engineering the level structure of a giant
  artificial atom in waveguide quantum electrodynamics},}\ }\href
  {https://arxiv.org/abs/2003.14167} {\bibfield  {journal} {\bibinfo  {journal}
  {arXiv preprint arXiv:2003.14167}\ } (\bibinfo {year} {2020})}\BibitemShut
  {NoStop}%
\bibitem [{\citenamefont {Witthaut}\ and\ \citenamefont
  {Sorensen}(2010)}]{WitthautNJP10}%
  \BibitemOpen
  \bibfield  {author} {\bibinfo {author} {\bibfnamefont {D.}~\bibnamefont
  {Witthaut}}\ and\ \bibinfo {author} {\bibfnamefont {A.~S.}\ \bibnamefont
  {Sorensen}},\ }\bibfield  {title} {\enquote {\bibinfo {title} {{Photon
  scattering by a three-level emitter in a one-dimensional waveguide}},}\
  }\href {\doibase 10.1088/1367-2630/12/4/043052} {\bibfield  {journal}
  {\bibinfo  {journal} {New Journal of Physics}\ }\textbf {\bibinfo {volume}
  {12}},\ \bibinfo {pages} {43052} (\bibinfo {year} {2010})},\ \Eprint
  {http://arxiv.org/abs/1001.0975} {1001.0975} \BibitemShut {NoStop}%
\bibitem [{\citenamefont {Fang}\ \emph {et~al.}(2018)\citenamefont {Fang},
  \citenamefont {Ciccarello},\ and\ \citenamefont {Baranger}}]{FangNJP18}%
  \BibitemOpen
  \bibfield  {author} {\bibinfo {author} {\bibfnamefont {Yao Lung~L.}\
  \bibnamefont {Fang}}, \bibinfo {author} {\bibfnamefont {Francesco}\
  \bibnamefont {Ciccarello}}, \ and\ \bibinfo {author} {\bibfnamefont
  {Harold~U.}\ \bibnamefont {Baranger}},\ }\bibfield  {title} {\enquote
  {\bibinfo {title} {{Non-Markovian dynamics of a qubit due to single-photon
  scattering in a waveguide}},}\ }\href {\doibase 10.1088/1367-2630/aaba5d}
  {\bibfield  {journal} {\bibinfo  {journal} {New Journal of Physics}\ }\textbf
  {\bibinfo {volume} {20}},\ \bibinfo {pages} {43035} (\bibinfo {year}
  {2018})}\BibitemShut {NoStop}%
\bibitem [{\citenamefont {Hoi}\ \emph {et~al.}(2015)\citenamefont {Hoi},
  \citenamefont {Kockum}, \citenamefont {Tornberg}, \citenamefont
  {Pourkabirian}, \citenamefont {Johansson}, \citenamefont {Delsing},\ and\
  \citenamefont {Wilson}}]{HoiNatPhy15}%
  \BibitemOpen
  \bibfield  {author} {\bibinfo {author} {\bibfnamefont {I.~C.}\ \bibnamefont
  {Hoi}}, \bibinfo {author} {\bibfnamefont {A.~F.}\ \bibnamefont {Kockum}},
  \bibinfo {author} {\bibfnamefont {L.}~\bibnamefont {Tornberg}}, \bibinfo
  {author} {\bibfnamefont {A.}~\bibnamefont {Pourkabirian}}, \bibinfo {author}
  {\bibfnamefont {G.}~\bibnamefont {Johansson}}, \bibinfo {author}
  {\bibfnamefont {P.}~\bibnamefont {Delsing}}, \ and\ \bibinfo {author}
  {\bibfnamefont {C.~M.}\ \bibnamefont {Wilson}},\ }\bibfield  {title}
  {\enquote {\bibinfo {title} {{Probing the quantum vacuum with an artificial
  atom in front of a mirror}},}\ }\href {\doibase 10.1038/nphys3484} {\bibfield
   {journal} {\bibinfo  {journal} {Nature Physics}\ }\textbf {\bibinfo {volume}
  {11}},\ \bibinfo {pages} {1045--1049} (\bibinfo {year} {2015})},\ \Eprint
  {http://arxiv.org/abs/1410.8840} {1410.8840} \BibitemShut {NoStop}%
\bibitem [{\citenamefont {Combes}\ \emph {et~al.}(2017)\citenamefont {Combes},
  \citenamefont {Kerckhoff},\ and\ \citenamefont {Sarovar}}]{CombesAdvPhyX17}%
  \BibitemOpen
  \bibfield  {author} {\bibinfo {author} {\bibfnamefont {Joshua}\ \bibnamefont
  {Combes}}, \bibinfo {author} {\bibfnamefont {Joseph}\ \bibnamefont
  {Kerckhoff}}, \ and\ \bibinfo {author} {\bibfnamefont {Mohan}\ \bibnamefont
  {Sarovar}},\ }\bibfield  {title} {\enquote {\bibinfo {title} {{The SLH
  framework for modeling quantum input-output networks}},}\ }\href {\doibase
  10.1080/23746149.2017.1343097} {\bibfield  {journal} {\bibinfo  {journal}
  {Advances in Physics: X}\ }\textbf {\bibinfo {volume} {2}},\ \bibinfo {pages}
  {784--888} (\bibinfo {year} {2017})}\BibitemShut {NoStop}%
\bibitem [{\citenamefont {Lalumi{\`{e}}re}\ \emph {et~al.}(2013)\citenamefont
  {Lalumi{\`{e}}re}, \citenamefont {Sanders}, \citenamefont {{Van Loo}},
  \citenamefont {Fedorov}, \citenamefont {Wallraff},\ and\ \citenamefont
  {Blais}}]{LalumierePRA13}%
  \BibitemOpen
  \bibfield  {author} {\bibinfo {author} {\bibfnamefont {Kevin}\ \bibnamefont
  {Lalumi{\`{e}}re}}, \bibinfo {author} {\bibfnamefont {Barry~C.}\ \bibnamefont
  {Sanders}}, \bibinfo {author} {\bibfnamefont {A.~F.}\ \bibnamefont {{Van
  Loo}}}, \bibinfo {author} {\bibfnamefont {A.}~\bibnamefont {Fedorov}},
  \bibinfo {author} {\bibfnamefont {A.}~\bibnamefont {Wallraff}}, \ and\
  \bibinfo {author} {\bibfnamefont {A.}~\bibnamefont {Blais}},\ }\bibfield
  {title} {\enquote {\bibinfo {title} {{Input-output theory for waveguide QED
  with an ensemble of inhomogeneous atoms}},}\ }\href {\doibase
  10.1103/PhysRevA.88.043806} {\bibfield  {journal} {\bibinfo  {journal}
  {Physical Review A - Atomic, Molecular, and Optical Physics}\ }\textbf
  {\bibinfo {volume} {88}},\ \bibinfo {pages} {43806} (\bibinfo {year}
  {2013})}\BibitemShut {NoStop}%
\bibitem [{\citenamefont {Ernst}\ \emph {et~al.}(1987)\citenamefont {Ernst},
  \citenamefont {Bodenhausen}, \citenamefont {Wokaun} \emph
  {et~al.}}]{ernst1987principles}%
  \BibitemOpen
  \bibfield  {author} {\bibinfo {author} {\bibfnamefont {Richard~R}\
  \bibnamefont {Ernst}}, \bibinfo {author} {\bibfnamefont {Geoffrey}\
  \bibnamefont {Bodenhausen}}, \bibinfo {author} {\bibfnamefont {Alexander}\
  \bibnamefont {Wokaun}},  \emph {et~al.},\ }\href@noop {} {\emph {\bibinfo
  {title} {Principles of nuclear magnetic resonance in one and two
  dimensions}}},\ Vol.~\bibinfo {volume} {14}\ (\bibinfo  {publisher}
  {Clarendon press Oxford},\ \bibinfo {year} {1987})\BibitemShut {NoStop}%
\bibitem [{\citenamefont {Vandersypen}\ and\ \citenamefont
  {Chuang}(2005)}]{ChuangRMP05}%
  \BibitemOpen
  \bibfield  {author} {\bibinfo {author} {\bibfnamefont {L.~M.~K.}\
  \bibnamefont {Vandersypen}}\ and\ \bibinfo {author} {\bibfnamefont {I.~L.}\
  \bibnamefont {Chuang}},\ }\bibfield  {title} {\enquote {\bibinfo {title} {Nmr
  techniques for quantum control and computation},}\ }\href {\doibase
  10.1103/RevModPhys.76.1037} {\bibfield  {journal} {\bibinfo  {journal} {Rev.
  Mod. Phys.}\ }\textbf {\bibinfo {volume} {76}},\ \bibinfo {pages}
  {1037--1069} (\bibinfo {year} {2005})}\BibitemShut {NoStop}%
\bibitem [{\citenamefont {James}\ and\ \citenamefont
  {Jerke}(2007)}]{JamesJCan07}%
  \BibitemOpen
  \bibfield  {author} {\bibinfo {author} {\bibfnamefont {D~F}\ \bibnamefont
  {James}}\ and\ \bibinfo {author} {\bibfnamefont {J}~\bibnamefont {Jerke}},\
  }\bibfield  {title} {\enquote {\bibinfo {title} {Effective hamiltonian theory
  and its applications in quantum information},}\ }\href {\doibase
  10.1139/p07-060} {\bibfield  {journal} {\bibinfo  {journal} {Canadian Journal
  of Physics}\ }\textbf {\bibinfo {volume} {85}},\ \bibinfo {pages} {625--632}
  (\bibinfo {year} {2007})}\BibitemShut {NoStop}%
\bibitem [{\citenamefont {Gamel}\ and\ \citenamefont
  {James}(2010)}]{JamesPRA10}%
  \BibitemOpen
  \bibfield  {author} {\bibinfo {author} {\bibfnamefont {Omar}\ \bibnamefont
  {Gamel}}\ and\ \bibinfo {author} {\bibfnamefont {Daniel F.~V.}\ \bibnamefont
  {James}},\ }\bibfield  {title} {\enquote {\bibinfo {title} {Time-averaged
  quantum dynamics and the validity of the effective hamiltonian model},}\
  }\href {\doibase 10.1103/PhysRevA.82.052106} {\bibfield  {journal} {\bibinfo
  {journal} {Phys. Rev. A}\ }\textbf {\bibinfo {volume} {82}},\ \bibinfo
  {pages} {052106} (\bibinfo {year} {2010})}\BibitemShut {NoStop}%
\bibitem [{\citenamefont {Di~Stefano}\ \emph {et~al.}(2016)\citenamefont
  {Di~Stefano}, \citenamefont {Paladino}, \citenamefont {Pope},\ and\
  \citenamefont {Falci}}]{FalciPRA16}%
  \BibitemOpen
  \bibfield  {author} {\bibinfo {author} {\bibfnamefont {P.~G.}\ \bibnamefont
  {Di~Stefano}}, \bibinfo {author} {\bibfnamefont {E.}~\bibnamefont
  {Paladino}}, \bibinfo {author} {\bibfnamefont {T.~J.}\ \bibnamefont {Pope}},
  \ and\ \bibinfo {author} {\bibfnamefont {G.}~\bibnamefont {Falci}},\
  }\bibfield  {title} {\enquote {\bibinfo {title} {Coherent manipulation of
  noise-protected superconducting artificial atoms in the lambda scheme},}\
  }\href {\doibase 10.1103/PhysRevA.93.051801} {\bibfield  {journal} {\bibinfo
  {journal} {Phys. Rev. A}\ }\textbf {\bibinfo {volume} {93}},\ \bibinfo
  {pages} {051801} (\bibinfo {year} {2016})}\BibitemShut {NoStop}%
\bibitem [{\citenamefont {Magnus}(1954)}]{Magnus1954}%
  \BibitemOpen
  \bibfield  {author} {\bibinfo {author} {\bibfnamefont {Wilhelm}\ \bibnamefont
  {Magnus}},\ }\bibfield  {title} {\enquote {\bibinfo {title} {On the
  exponential solution of differential equations for a linear operator},}\
  }\href@noop {} {\bibfield  {journal} {\bibinfo  {journal} {Communications on
  pure and applied mathematics}\ }\textbf {\bibinfo {volume} {7}},\ \bibinfo
  {pages} {649--673} (\bibinfo {year} {1954})}\BibitemShut {NoStop}%
\bibitem [{\citenamefont {Haroche}\ and\ \citenamefont
  {Raimond}(2006)}]{harocheExploring2006}%
  \BibitemOpen
  \bibfield  {author} {\bibinfo {author} {\bibfnamefont {Serge}\ \bibnamefont
  {Haroche}}\ and\ \bibinfo {author} {\bibfnamefont {Jean-Michel}\ \bibnamefont
  {Raimond}},\ }\href
  {http://www.oxfordscholarship.com/view/10.1093/acprof:oso/9780198509141.001.0001/acprof-9780198509141}
  {\emph {\bibinfo {title} {Exploring the Quantum : Atoms, Cavities , and
  Photons}}}\ (\bibinfo  {publisher} {Oxford University Press},\ \bibinfo
  {year} {2006})\BibitemShut {NoStop}%
\bibitem [{\citenamefont {Reiter}\ and\ \citenamefont
  {S\o{}rensen}(2012)}]{SorensenPRA11}%
  \BibitemOpen
  \bibfield  {author} {\bibinfo {author} {\bibfnamefont {Florentin}\
  \bibnamefont {Reiter}}\ and\ \bibinfo {author} {\bibfnamefont {Anders~S.}\
  \bibnamefont {S\o{}rensen}},\ }\bibfield  {title} {\enquote {\bibinfo {title}
  {Effective operator formalism for open quantum systems},}\ }\href {\doibase
  10.1103/PhysRevA.85.032111} {\bibfield  {journal} {\bibinfo  {journal} {Phys.
  Rev. A}\ }\textbf {\bibinfo {volume} {85}},\ \bibinfo {pages} {032111}
  (\bibinfo {year} {2012})}\BibitemShut {NoStop}%
\bibitem [{\citenamefont {Peters}(2012)}]{HarocheRaimondBook}%
  \BibitemOpen
  \bibfield  {author} {\bibinfo {author} {\bibfnamefont {Lauren}\ \bibnamefont
  {Peters}},\ }\href {\doibase 10.4103/0974-6102.97661} {\emph {\bibinfo
  {title} {Young Scientists Journal}}},\ Vol.~\bibinfo {volume} {5}\ (\bibinfo
  {publisher} {Oxford Univ. Press},\ \bibinfo {address} {Oxford, UK},\ \bibinfo
  {year} {2012})\ p.~\bibinfo {pages} {11}\BibitemShut {NoStop}%
\bibitem [{\citenamefont {Roy}\ \emph {et~al.}(2017)\citenamefont {Roy},
  \citenamefont {Wilson},\ and\ \citenamefont {Firstenberg}}]{RoyRMP17}%
  \BibitemOpen
  \bibfield  {author} {\bibinfo {author} {\bibfnamefont {Dibyendu}\
  \bibnamefont {Roy}}, \bibinfo {author} {\bibfnamefont {C.~M.}\ \bibnamefont
  {Wilson}}, \ and\ \bibinfo {author} {\bibfnamefont {Ofer}\ \bibnamefont
  {Firstenberg}},\ }\bibfield  {title} {\enquote {\bibinfo {title}
  {{Colloquium: Strongly interacting photons in one-dimensional continuum}},}\
  }\href {\doibase 10.1103/RevModPhys.89.021001} {\bibfield  {journal}
  {\bibinfo  {journal} {Reviews of Modern Physics}\ }\textbf {\bibinfo {volume}
  {89}},\ \bibinfo {pages} {21001} (\bibinfo {year} {2017})}\BibitemShut
  {NoStop}%
\bibitem [{\citenamefont {Liao}\ \emph {et~al.}(2016)\citenamefont {Liao},
  \citenamefont {Zeng}, \citenamefont {Nha},\ and\ \citenamefont
  {Zubairy}}]{LiaoPhyScr16}%
  \BibitemOpen
  \bibfield  {author} {\bibinfo {author} {\bibfnamefont {Zeyang}\ \bibnamefont
  {Liao}}, \bibinfo {author} {\bibfnamefont {Xiaodong}\ \bibnamefont {Zeng}},
  \bibinfo {author} {\bibfnamefont {Hyunchul}\ \bibnamefont {Nha}}, \ and\
  \bibinfo {author} {\bibfnamefont {M.~Suhail}\ \bibnamefont {Zubairy}},\
  }\bibfield  {title} {\enquote {\bibinfo {title} {{Photon transport in a
  one-dimensional nanophotonic waveguide QED system}},}\ }\href {\doibase
  10.1088/0031-8949/91/6/063004} {\bibfield  {journal} {\bibinfo  {journal}
  {Physica Scripta}\ }\textbf {\bibinfo {volume} {91}},\ \bibinfo {pages}
  {63004} (\bibinfo {year} {2016})}\BibitemShut {NoStop}%
\bibitem [{\citenamefont {Gu}\ \emph {et~al.}(2017)\citenamefont {Gu},
  \citenamefont {Kockum}, \citenamefont {Miranowicz}, \citenamefont {xi~Liu},\
  and\ \citenamefont {Nori}}]{GuarXiv17}%
  \BibitemOpen
  \bibfield  {author} {\bibinfo {author} {\bibfnamefont {Xiu}\ \bibnamefont
  {Gu}}, \bibinfo {author} {\bibfnamefont {Anton~Frisk}\ \bibnamefont
  {Kockum}}, \bibinfo {author} {\bibfnamefont {Adam}\ \bibnamefont
  {Miranowicz}}, \bibinfo {author} {\bibfnamefont {Yu}~\bibnamefont {xi~Liu}},
  \ and\ \bibinfo {author} {\bibfnamefont {Franco}\ \bibnamefont {Nori}},\
  }\bibfield  {title} {\enquote {\bibinfo {title} {{Microwave photonics with
  superconducting quantum circuits}},}\ }\href {\doibase
  10.1016/j.physrep.2017.10.002} {\bibfield  {journal} {\bibinfo  {journal}
  {Physics Reports}\ }\textbf {\bibinfo {volume} {718-719}},\ \bibinfo {pages}
  {1--102} (\bibinfo {year} {2017})}\BibitemShut {NoStop}%
\bibitem [{not({\natexlab{a}})}]{nota-trivial}%
  \BibitemOpen
  \href@noop {} {}\bibinfo {howpublished} {So long as we deal with giant atoms
  having two coupling points each, a trivial Hamiltonian is always identically
  zero. Later on, in Section \ref{sec-multi}, we will see that a $H_{\rm eff}$
  can be non-zero but feature zero inter-atomic couplings, this being still
  trivial when the goal is coupling the atoms.} ({\natexlab{a}})\BibitemShut
  {NoStop}%
\bibitem [{\citenamefont {Cilluffo}\ \emph {et~al.}(2020)\citenamefont
  {Cilluffo}, \citenamefont {Carollo}, \citenamefont {Lorenzo}, \citenamefont
  {Gross}, \citenamefont {Palma},\ and\ \citenamefont {Ciccarello}}]{CP}%
  \BibitemOpen
  \bibfield  {author} {\bibinfo {author} {\bibfnamefont {Dario}\ \bibnamefont
  {Cilluffo}}, \bibinfo {author} {\bibfnamefont {Angelo}\ \bibnamefont
  {Carollo}}, \bibinfo {author} {\bibfnamefont {Salvatore}\ \bibnamefont
  {Lorenzo}}, \bibinfo {author} {\bibfnamefont {Jonathan~A}\ \bibnamefont
  {Gross}}, \bibinfo {author} {\bibfnamefont {G~Massimo}\ \bibnamefont
  {Palma}}, \ and\ \bibinfo {author} {\bibfnamefont {Francesco}\ \bibnamefont
  {Ciccarello}},\ }\bibfield  {title} {\enquote {\bibinfo {title} {Collisional
  picture of quantum optics with giant emitters},}\ }\href
  {https://arxiv.org/abs/2006.08631} {\bibfield  {journal} {\bibinfo  {journal}
  {Phys. Rev. Research}\ }\textbf {\bibinfo {volume} {2}},\ \bibinfo {pages}
  {43070} (\bibinfo {year} {2020})}\BibitemShut {NoStop}%
\bibitem [{\citenamefont {Ciccarello}(2017)}]{ciccarelloCollision2017}%
  \BibitemOpen
  \bibfield  {author} {\bibinfo {author} {\bibfnamefont {Francesco}\
  \bibnamefont {Ciccarello}},\ }\bibfield  {title} {\enquote {\bibinfo {title}
  {{Collision models in quantum optics}},}\ }\href
  {https://doi.org/10.1515/qmetro-2017-0007} {\bibfield  {journal} {\bibinfo
  {journal} {Quantum Measurements and Quantum Metrology}\ }\textbf {\bibinfo
  {volume} {4}},\ \bibinfo {pages} {53} (\bibinfo {year} {2017})}\BibitemShut
  {NoStop}%
\bibitem [{\citenamefont {Brun}(2002)}]{brunSimple2002}%
  \BibitemOpen
  \bibfield  {author} {\bibinfo {author} {\bibfnamefont {Todd~A.}\ \bibnamefont
  {Brun}},\ }\bibfield  {title} {\enquote {\bibinfo {title} {{A simple model of
  quantum trajectories}},}\ }\href {\doibase 10.1119/1.1475328} {\bibfield
  {journal} {\bibinfo  {journal} {American Journal of Physics}\ }\textbf
  {\bibinfo {volume} {70}},\ \bibinfo {pages} {719--737} (\bibinfo {year}
  {2002})}\BibitemShut {NoStop}%
\bibitem [{\citenamefont {Altamirano}\ \emph {et~al.}(2017)\citenamefont
  {Altamirano}, \citenamefont {Corona-Ugalde}, \citenamefont {Mann},\ and\
  \citenamefont {Zych}}]{altamiranoUnitarity2017}%
  \BibitemOpen
  \bibfield  {author} {\bibinfo {author} {\bibfnamefont {Natacha}\ \bibnamefont
  {Altamirano}}, \bibinfo {author} {\bibfnamefont {Paulina}\ \bibnamefont
  {Corona-Ugalde}}, \bibinfo {author} {\bibfnamefont {Robert~B.}\ \bibnamefont
  {Mann}}, \ and\ \bibinfo {author} {\bibfnamefont {Magdalena}\ \bibnamefont
  {Zych}},\ }\bibfield  {title} {\enquote {\bibinfo {title} {{Unitarity,
  feedback, interactions - Dynamics emergent from repeated measurements}},}\
  }\href {\doibase 10.1088/1367-2630/aa551b} {\bibfield  {journal} {\bibinfo
  {journal} {New Journal of Physics}\ }\textbf {\bibinfo {volume} {19}},\
  \bibinfo {pages} {13035} (\bibinfo {year} {2017})}\BibitemShut {NoStop}%
\bibitem [{\citenamefont {Lorenzo}\ \emph {et~al.}(2017)\citenamefont
  {Lorenzo}, \citenamefont {Ciccarello},\ and\ \citenamefont
  {Palma}}]{lorenzoComposite2017}%
  \BibitemOpen
  \bibfield  {author} {\bibinfo {author} {\bibfnamefont {Salvatore}\
  \bibnamefont {Lorenzo}}, \bibinfo {author} {\bibfnamefont {Francesco}\
  \bibnamefont {Ciccarello}}, \ and\ \bibinfo {author} {\bibfnamefont
  {G.~Massimo}\ \bibnamefont {Palma}},\ }\bibfield  {title} {\enquote {\bibinfo
  {title} {{Composite quantum collision models}},}\ }\href {\doibase
  10.1103/PhysRevA.96.032107} {\bibfield  {journal} {\bibinfo  {journal}
  {Physical Review A}\ }\textbf {\bibinfo {volume} {96}},\ \bibinfo {pages}
  {32107} (\bibinfo {year} {2017})},\ \Eprint {http://arxiv.org/abs/1705.03215}
  {1705.03215} \BibitemShut {NoStop}%
\bibitem [{\citenamefont {Gross}\ \emph {et~al.}(2018)\citenamefont {Gross},
  \citenamefont {Caves}, \citenamefont {Milburn},\ and\ \citenamefont
  {Combes}}]{grossQubit2018}%
  \BibitemOpen
  \bibfield  {author} {\bibinfo {author} {\bibfnamefont {Jonathan~A.}\
  \bibnamefont {Gross}}, \bibinfo {author} {\bibfnamefont {Carlton~M.}\
  \bibnamefont {Caves}}, \bibinfo {author} {\bibfnamefont {Gerard~J.}\
  \bibnamefont {Milburn}}, \ and\ \bibinfo {author} {\bibfnamefont {Joshua}\
  \bibnamefont {Combes}},\ }\bibfield  {title} {\enquote {\bibinfo {title}
  {{Qubit models of weak continuous measurements: Markovian conditional and
  open-system dynamics}},}\ }\href {\doibase 10.1088/2058-9565/aaa39f}
  {\bibfield  {journal} {\bibinfo  {journal} {Quantum Science and Technology}\
  }\textbf {\bibinfo {volume} {3}},\ \bibinfo {pages} {024005} (\bibinfo {year}
  {2018})}\BibitemShut {NoStop}%
\bibitem [{\citenamefont {Giovannetti}\ and\ \citenamefont
  {Palma}(2012{\natexlab{a}})}]{giovannettiMaster2012a}%
  \BibitemOpen
  \bibfield  {author} {\bibinfo {author} {\bibfnamefont {V.}~\bibnamefont
  {Giovannetti}}\ and\ \bibinfo {author} {\bibfnamefont {G.~M.}\ \bibnamefont
  {Palma}},\ }\bibfield  {title} {\enquote {\bibinfo {title} {{Master equations
  for correlated quantum channels}},}\ }\href {\doibase
  10.1103/PhysRevLett.108.040401} {\bibfield  {journal} {\bibinfo  {journal}
  {Physical Review Letters}\ }\textbf {\bibinfo {volume} {108}},\ \bibinfo
  {pages} {40401} (\bibinfo {year} {2012}{\natexlab{a}})},\ \Eprint
  {http://arxiv.org/abs/1105.4506} {arXiv:1105.4506} \BibitemShut {NoStop}%
\bibitem [{\citenamefont {Giovannetti}\ and\ \citenamefont
  {Palma}(2012{\natexlab{b}})}]{giovannettiMaster2012}%
  \BibitemOpen
  \bibfield  {author} {\bibinfo {author} {\bibfnamefont {V.}~\bibnamefont
  {Giovannetti}}\ and\ \bibinfo {author} {\bibfnamefont {G.~M.}\ \bibnamefont
  {Palma}},\ }\bibfield  {title} {\enquote {\bibinfo {title} {{Master equation
  for cascade quantum channels: A collisional approach}},}\ }\href {\doibase
  10.1088/0953-4075/45/15/154003} {\bibfield  {journal} {\bibinfo  {journal}
  {Journal of Physics B: Atomic, Molecular and Optical Physics}\ }\textbf
  {\bibinfo {volume} {45}},\ \bibinfo {pages} {154003} (\bibinfo {year}
  {2012}{\natexlab{b}})}\BibitemShut {NoStop}%
\bibitem [{\citenamefont {Lorenzo}\ \emph {et~al.}(2015)\citenamefont
  {Lorenzo}, \citenamefont {Farace}, \citenamefont {Ciccarello}, \citenamefont
  {Palma},\ and\ \citenamefont {Giovannetti}}]{LorenzoFlux}%
  \BibitemOpen
  \bibfield  {author} {\bibinfo {author} {\bibfnamefont {Salvatore}\
  \bibnamefont {Lorenzo}}, \bibinfo {author} {\bibfnamefont {Alessandro}\
  \bibnamefont {Farace}}, \bibinfo {author} {\bibfnamefont {Francesco}\
  \bibnamefont {Ciccarello}}, \bibinfo {author} {\bibfnamefont {G.~Massimo}\
  \bibnamefont {Palma}}, \ and\ \bibinfo {author} {\bibfnamefont {Vittorio}\
  \bibnamefont {Giovannetti}},\ }\bibfield  {title} {\enquote {\bibinfo {title}
  {Heat flux and quantum correlations in dissipative cascaded systems},}\
  }\href {\doibase 10.1103/PhysRevA.91.022121} {\bibfield  {journal} {\bibinfo
  {journal} {Phys. Rev. A}\ }\textbf {\bibinfo {volume} {91}},\ \bibinfo
  {pages} {022121} (\bibinfo {year} {2015})}\BibitemShut {NoStop}%
\bibitem [{\citenamefont {Lodahl}\ \emph {et~al.}(2017)\citenamefont {Lodahl},
  \citenamefont {Mahmoodian}, \citenamefont {Stobbe}, \citenamefont
  {Rauschenbeutel}, \citenamefont {Schneeweiss}, \citenamefont {Volz},
  \citenamefont {Pichler},\ and\ \citenamefont
  {Zoller}}]{LodahlReviewNature17}%
  \BibitemOpen
  \bibfield  {author} {\bibinfo {author} {\bibfnamefont {Peter}\ \bibnamefont
  {Lodahl}}, \bibinfo {author} {\bibfnamefont {Sahand}\ \bibnamefont
  {Mahmoodian}}, \bibinfo {author} {\bibfnamefont {S{\o}ren}\ \bibnamefont
  {Stobbe}}, \bibinfo {author} {\bibfnamefont {Arno}\ \bibnamefont
  {Rauschenbeutel}}, \bibinfo {author} {\bibfnamefont {Philipp}\ \bibnamefont
  {Schneeweiss}}, \bibinfo {author} {\bibfnamefont {J{\"{u}}rgen}\ \bibnamefont
  {Volz}}, \bibinfo {author} {\bibfnamefont {Hannes}\ \bibnamefont {Pichler}},
  \ and\ \bibinfo {author} {\bibfnamefont {Peter}\ \bibnamefont {Zoller}},\
  }\bibfield  {title} {\enquote {\bibinfo {title} {{Chiral quantum optics}},}\
  }\href {\doibase 10.1038/nature21037} {\bibfield  {journal} {\bibinfo
  {journal} {Nature}\ }\textbf {\bibinfo {volume} {541}},\ \bibinfo {pages}
  {473--480} (\bibinfo {year} {2017})}\BibitemShut {NoStop}%
\bibitem [{\citenamefont {Wang}(2012)}]{Nielsen00}%
  \BibitemOpen
  \bibfield  {author} {\bibinfo {author} {\bibfnamefont {Yazhen}\ \bibnamefont
  {Wang}},\ }\href {\doibase 10.1214/11-STS378} {\emph {\bibinfo {title}
  {Statistical Science}}},\ Vol.~\bibinfo {volume} {27}\ (\bibinfo  {publisher}
  {Cambridge University Express},\ \bibinfo {address} {Cambridge},\ \bibinfo
  {year} {2012})\ pp.\ \bibinfo {pages} {373--394},\ \Eprint
  {http://arxiv.org/abs/0512125} {0512125 [quant-ph]} \BibitemShut {NoStop}%
\bibitem [{not({\natexlab{b}})}]{note1}%
  \BibitemOpen
  \href@noop {} {}\bibinfo {howpublished} {In this case, before the collision,
  the time bin is in the vacuum state $\varrho_n=\ket{0}_n\!\bra{0}$. Using
  \eqref{Vnnu} and \eqref{deco}, one can check that in $U_n\, \varrho_n
  \ket{0}_n\!\bra{0}U_n^\dag$ terms $\sim \ket{k'}_n\!\bra{k}$ with $k\,({\rm
  \,or}\,k')=2,3...$ are at least of order $\sim \Delta t^{3/2}$. Thus the time
  bin behaves as a qubit with computational basis $\{\ket{0}_n,\ket{1}_n\}$.}
  ({\natexlab{b}})\BibitemShut {NoStop}%
\bibitem [{\citenamefont {Viola}\ \emph {et~al.}(1999)\citenamefont {Viola},
  \citenamefont {Knill},\ and\ \citenamefont {Lloyd}}]{ViolaPRL1999}%
  \BibitemOpen
  \bibfield  {author} {\bibinfo {author} {\bibfnamefont {Lorenza}\ \bibnamefont
  {Viola}}, \bibinfo {author} {\bibfnamefont {Emanuel}\ \bibnamefont {Knill}},
  \ and\ \bibinfo {author} {\bibfnamefont {Seth}\ \bibnamefont {Lloyd}},\
  }\bibfield  {title} {\enquote {\bibinfo {title} {{Dynamical Decoupling of
  Open Quantum Systems}},}\ }\href {\doibase 10.1103/PhysRevLett.82.2417}
  {\bibfield  {journal} {\bibinfo  {journal} {Phys. Rev. Lett.}\ }\textbf
  {\bibinfo {volume} {82}},\ \bibinfo {pages} {2417--2421} (\bibinfo {year}
  {1999})},\ \Eprint {http://arxiv.org/abs/9809071} {9809071} \BibitemShut
  {NoStop}%
\bibitem [{\citenamefont {S\o{}rensen}\ and\ \citenamefont
  {M\o{}lmer}(2002)}]{MolmerPRA02}%
  \BibitemOpen
  \bibfield  {author} {\bibinfo {author} {\bibfnamefont {Anders~S\o{}ndberg}\
  \bibnamefont {S\o{}rensen}}\ and\ \bibinfo {author} {\bibfnamefont {Klaus}\
  \bibnamefont {M\o{}lmer}},\ }\bibfield  {title} {\enquote {\bibinfo {title}
  {Entangling atoms in bad cavities},}\ }\href {\doibase
  10.1103/PhysRevA.66.022314} {\bibfield  {journal} {\bibinfo  {journal} {Phys.
  Rev. A}\ }\textbf {\bibinfo {volume} {66}},\ \bibinfo {pages} {022314}
  (\bibinfo {year} {2002})}\BibitemShut {NoStop}%
\bibitem [{\citenamefont {Gonz\'alez-Tudela}\ \emph {et~al.}(2019)\citenamefont
  {Gonz\'alez-Tudela}, \citenamefont {Mu\~noz},\ and\ \citenamefont
  {Cirac}}]{TudelaGiant19}%
  \BibitemOpen
  \bibfield  {author} {\bibinfo {author} {\bibfnamefont {A.}~\bibnamefont
  {Gonz\'alez-Tudela}}, \bibinfo {author} {\bibfnamefont {C.~S\'anchez}\
  \bibnamefont {Mu\~noz}}, \ and\ \bibinfo {author} {\bibfnamefont {J.~I.}\
  \bibnamefont {Cirac}},\ }\bibfield  {title} {\enquote {\bibinfo {title}
  {Engineering and harnessing giant atoms in high-dimensional baths: A proposal
  for implementation with cold atoms},}\ }\href {\doibase
  10.1103/PhysRevLett.122.203603} {\bibfield  {journal} {\bibinfo  {journal}
  {Phys. Rev. Lett.}\ }\textbf {\bibinfo {volume} {122}},\ \bibinfo {pages}
  {203603} (\bibinfo {year} {2019})}\BibitemShut {NoStop}%
\bibitem [{\citenamefont {Sutton}\ and\ \citenamefont
  {Datta}(2015)}]{suttonManipulating2015}%
  \BibitemOpen
  \bibfield  {author} {\bibinfo {author} {\bibfnamefont {Brian}\ \bibnamefont
  {Sutton}}\ and\ \bibinfo {author} {\bibfnamefont {Supriyo}\ \bibnamefont
  {Datta}},\ }\bibfield  {title} {\enquote {\bibinfo {title} {{Manipulating
  quantum information with spin torque}},}\ }\href {\doibase 10.1038/srep17912}
  {\bibfield  {journal} {\bibinfo  {journal} {Scientific Reports}\ }\textbf
  {\bibinfo {volume} {5}},\ \bibinfo {pages} {17912} (\bibinfo {year}
  {2015})}\BibitemShut {NoStop}%
\bibitem [{\citenamefont {Layden}\ \emph {et~al.}(2016)\citenamefont {Layden},
  \citenamefont {Mart{\'{i}}n-Mart{\'{i}}nez},\ and\ \citenamefont
  {Kempf}}]{laydenUniversal2016}%
  \BibitemOpen
  \bibfield  {author} {\bibinfo {author} {\bibfnamefont {David}\ \bibnamefont
  {Layden}}, \bibinfo {author} {\bibfnamefont {Eduardo}\ \bibnamefont
  {Mart{\'{i}}n-Mart{\'{i}}nez}}, \ and\ \bibinfo {author} {\bibfnamefont
  {Achim}\ \bibnamefont {Kempf}},\ }\bibfield  {title} {\enquote {\bibinfo
  {title} {{Universal scheme for indirect quantum control}},}\ }\href {\doibase
  10.1103/PhysRevA.93.040301} {\bibfield  {journal} {\bibinfo  {journal}
  {Physical Review A}\ }\textbf {\bibinfo {volume} {93}},\ \bibinfo {pages}
  {40301} (\bibinfo {year} {2016})},\ \Eprint {http://arxiv.org/abs/1506.06749}
  {1506.06749} \BibitemShut {NoStop}%
\bibitem [{\citenamefont {\c{C}akmak}\ \emph {et~al.}(2019)\citenamefont
  {\c{C}akmak}, \citenamefont {Campbell}, \citenamefont {Vacchini},
  \citenamefont {M\"ustecapl\ifmmode \imath \else \i
  \fi{}o\ifmmode~\breve{g}\else \u{g}\fi{}lu},\ and\ \citenamefont
  {Paternostro}}]{CampbellPRA19}%
  \BibitemOpen
  \bibfield  {author} {\bibinfo {author} {\bibfnamefont {Barl\c{s}}\
  \bibnamefont {\c{C}akmak}}, \bibinfo {author} {\bibfnamefont {Steve}\
  \bibnamefont {Campbell}}, \bibinfo {author} {\bibfnamefont {Bassano}\
  \bibnamefont {Vacchini}}, \bibinfo {author} {\bibfnamefont {\"Ozg\"ur~E.}\
  \bibnamefont {M\"ustecapl\ifmmode \imath \else \i
  \fi{}o\ifmmode~\breve{g}\else \u{g}\fi{}lu}}, \ and\ \bibinfo {author}
  {\bibfnamefont {Mauro}\ \bibnamefont {Paternostro}},\ }\bibfield  {title}
  {\enquote {\bibinfo {title} {Robust multipartite entanglement generation via
  a collision model},}\ }\href {\doibase 10.1103/PhysRevA.99.012319} {\bibfield
   {journal} {\bibinfo  {journal} {Phys. Rev. A}\ }\textbf {\bibinfo {volume}
  {99}},\ \bibinfo {pages} {012319} (\bibinfo {year} {2019})}\BibitemShut
  {NoStop}%
\end{thebibliography}%

\end{document}